\documentclass[a4paper,11pt]{article}
\pdfoutput=1 

\usepackage{jheppub} 

\usepackage[T1]{fontenc} 

\usepackage{bm}
\usepackage{JHEP_PR_dilasym}
\usepackage{subfigure}
\usepackage{caption}
\usepackage{lscape}
\usepackage{float}

\usepackage{preprintcover}
\PreprintCoverPaperTitle{Measurement of the charge asymmetry in dileptonic decays of top quark pairs in \pp collisions at $\rts=7$\,\tev\ using the \atlas detector}
\PreprintIdNumber{CERN-PH-EP-2014-233}  
\PreprintCoverAbstract{
A measurement of the \ttbarw (\ttbar) charge asymmetry is presented
using data corresponding to an integrated luminosity of \lumi of LHC \pp collisions at a
centre-of-mass energy of 7\,\tev\ collected by the \atlas detector.
Events with two charged leptons, at least two jets and large missing transverse momentum are selected.
Two observables are studied: \Acll based on the identified charged
leptons, and \Actt, based on the reconstructed \ttbar\ final state.
The asymmetries are measured to be
\begin{eqnarray*}
 \Acll & = & \Acllcomb,
\end{eqnarray*}
\begin{eqnarray*}
 \Actt & = & \Acttcomb.
\end{eqnarray*}
The measured values are in agreement with the Standard Model predictions.
}
\PreprintJournalName{JHEP}

\title{\boldmath Measurement of the charge asymmetry in dileptonic decays of top quark pairs in \pp collisions at $\rts=7$\,\tev\ using the \atlas detector}

\collaboration{The \atlas collaboration}

\abstract{A measurement of the \ttbarw (\ttbar) charge asymmetry is presented 
using data corresponding to an integrated luminosity of \lumi of LHC \pp collisions at a 
centre-of-mass energy of 7\,\tev\ collected by the \atlas detector. 
Events with two charged leptons, at least two jets and large missing transverse momentum are selected. 
Two observables are studied: \Acll based on the identified charged
leptons, and \Actt, based on the reconstructed \ttbar\ final state.  
The asymmetries are measured to be
\begin{eqnarray*}
 \Acll & = & \Acllcomb,
\end{eqnarray*}
\begin{eqnarray*}
 \Actt & = & \Acttcomb.
\end{eqnarray*}
The measured values are in agreement with the Standard Model predictions.}
 
\begin{document} 
\maketitle
\flushbottom

\section{Introduction}
\label{sec:Introduction}
The top quark is the heaviest elementary particle known to date. It was
discovered in 1995 at the Tevatron proton--antiproton (\ppbar) collider by the
CDF and D0 collaborations~\cite{TOP1,TOP2}. It is the only quark in the Standard
Model (SM) that decays before hadronization occurs, and the only quark with
Yukawa coupling to the Higgs boson close to unity. A precise study of top
quark properties could shed light on possible physics models beyond the
SM~\cite{Frederix:2007gi,Barger:2006hm,AguilarSaavedra:2010zi,Zhang:2010dr,Hill:1994hp,Cao:2004wd,Aguilar-Saavedra:2014nja}. 

This analysis uses a data set corresponding to an integrated luminosity of
\lumi{} of Large Hadron Collider (LHC)  proton--proton (\pp{}) collisions at a
centre-of-mass energy of 7\,\tev\ collected by the \atlas{} detector. It is
performed in the dilepton channel of the \ttbar{} pair decay, realized when both
$W$ bosons decay to a charged lepton and a neutrino. The measured observables
are the lepton-based charge asymmetry \Acll and the \ttbar{} charge asymmetry \Actt{}.
The observable \Acll is defined as an asymmetry between positively and
negatively charged leptons (electrons and muons) in the dilepton decays of the
\ttbar{} pairs, 
\begin{equation}
\label{eq:ac_lep}
\Acll = \frac{N(\Delta |\eta| >0) - N(\Delta |\eta| <0)}{N(\Delta |\eta|
      >0) + N(\Delta |\eta| <0)}, 
\end{equation}
where 
\begin{equation}
\Delta |\eta| = |\eta_{\ell^{+}}|-|\eta_{\ell^{-}}|,
\end{equation}
$\eta_{\ell^{+}}$ ($\eta_{\ell^{-}}$) is the pseudorapidity\footnote{The pseudorapidity is defined in 
terms of the polar angle $\theta$ as $\eta=-\ln\tan(\theta/2)$.} 
of the positively (negatively) charged lepton and $N$ is the number of events
with positive or negative $\Delta |\eta|$.
While \Acll is defined in~\eqRef{eq:ac_lep} as an asymmetry between positively and
negatively charged lepton pseudorapidities, \Actt{} corresponds to the asymmetry in top quark and
antitop quark rapidities\footnote{The rapidity is defined as 
$y = \frac{1}{2} \ln\frac{E + p_z}{E - p_z}$ where $E$ is the energy of the
particle and $p_z$ is the component of the momentum along the LHC beam axis.},
\begin{equation}
\label{eq:ac}
\Actt = \frac{N(\Delta |y| >0) - N(\Delta |y| <0)}{N(\Delta |y|
  >0) + N(\Delta |y| <0)}, 
\end{equation}
where 
\begin{equation}
\Delta |y| = |y_{t}|-|y_{\bar{t}}|, 
\end{equation}
$y_{t}$ ($y_{\bar{t}}$) is the rapidity of the top (antitop) quark, and $N$ is the number of events with positive or
negative $\Delta |y|$.

In SM \ttbar{} production, the asymmetry is absent at leading-order (LO) in Quantum Chromodynamics (QCD) and is introduced by the
next-to-leading-order (NLO) QCD contributions to the \ttbar{} differential
cross-sections, which are odd with respect to the exchange of $t$ and $\tbar$. At
the LHC, the contributions to the asymmetries defined in \eqRef{eq:ac_lep} and
\eqRef{eq:ac} are predominantly from $\qqbar$-initiated \ttbar{} production,
and $qg$-initiated production also has a non-negligible contribution.
The $gg$-initiated processes are symmetric~\cite{Bernreuther:2012sx}.
The asymmetry predicted in the SM is slightly positive, implying that the top quark is 
preferentially emitted in the direction of the quark in the initial state. 
In $\qqbar$ interactions at the LHC, the quark is in most cases a valence quark whereas 
the antiquark is from the sea. The asymmetry translates to a higher boost along
the beam direction for the $t$-quark than for the $\tbar$-quark. The rapidity
distribution of the $t$ is thus slightly broader than the one of the $\tbar$.
      
The SM predictions of \Actt{} and \Acll{} computed at NLO in QCD and including electroweak corrections (NLO QCD+EW) are~\cite{Bernreuther:2012sx} 
\begin{equation}
\label{eq:Actttheory}
\Actt = \Actttheory,
\end{equation}
\begin{equation}
\label{eq:Aclltheory}
\Acll = \Aclltheory.
\end{equation}
These asymmetries are evaluated without acceptance cuts. The uncertainties
are due to scale variations, estimated by simultaneous variation of the
renormalization and factorization scale by a factor of half or two with respect
to the reference scale value, which is set to the top quark mass. 
Recent next-to-next-to-leading order (NNLO) calculations of the forward-backward asymmetry for the Tevatron suggest that varying these scales significantly underestimates the uncertainty due to higher order corrections~\cite{Czakon:2014xsa}, but no NNLO calculation has yet been published for \pp{} interactions at the LHC energies. There is however a recent calculation obtained with the Principle of Maximum Conformality~\cite{Wang:2014sua} which gives a consistent value of \Actt=\ActttheoPMC. The predicted value of \Acll{} is smaller than
the prediction for \Actt, since the directions of the leptons do not fully follow the
direction of the parent $t$ and $\tbar$ quarks. However, \Acll{} can be
measured more precisely, since it is determined without the need for a full
reconstruction of $t$ and $\tbar$ kinematics, which involves the use of jets and
missing transverse momentum that are reconstructed with less precision than the
kinematic variables of the leptons. The values
of \Acll{} and \Actt{} as well as their correlation can be sensitive to new
physics arising in top quark pair production~\cite{Frampton:1987dn,Ferrario:2008wm,AguilarSaavedra:2011vw,Falkowski:2012cu,Aguilar-Saavedra:2014kpa}. 

The asymmetry \Actt{} has been measured in the single-lepton decay channel by
the ATLAS~\cite{Aad:2013cea} and CMS~\cite{Chatrchyan:2012xv} collaborations at
$\rts=7$\,\tev{}. The CMS collaboration has reported measurements of \Acll{} and
\Actt{} in the dilepton decay channel at $\rts=7$\,\tev{}~\cite{Chatrchyan:2014yta}. 
The measured asymmetry values as well as 
those from a combination of ATLAS and CMS \Actt{} results in the single-lepton
decay channel~\cite{ATLAS-CONF-2014-012} are consistent with the SM predictions. 

At the Tevatron collider, \ttbar{} production has a forward-backward asymmetry with respect to the direction of the proton and antiproton beams. The asymmetry based on $t$ and $\tbar$ quarks, \Afbtt{}, is defined as 
\begin{equation}
\label{eq:afb}
\Afbtt = \frac{N(\Delta y >0) - N(\Delta y <0)}{N(\Delta y >0) + N(\Delta y <0)}, 
\end{equation}
where 
\begin{equation}
\Delta y = y_{t} - y_{\bar{t}}, 
\end{equation}
$y_{t}$ ($y_{\bar{t}}$) is the rapidity of the $t$ ($\tbar$) quark and $N$ is
the number of events with positive or negative $\Delta y$. An analogously defined
lepton-based forward-backward asymmetry in \ttbar{} production has been
studied as well. At the Tevatron, \ttbar{} events are predominantly produced by
\qqbar{} annihilation, thus the predicted asymmetries are typically larger than
at the LHC, where $gg$-initiated production dominates. The Tevatron experiments
have reported deviations of forward-backward asymmetries from the SM
predictions~\cite{CDF1,D01}, which have motivated a number of further asymmetry
measurements. Comparing the results with the latest NNLO calculations available
at the Tevatron~\cite{Czakon:2014xsa}, the deviations reported by the CDF
collaboration~\cite{CDF2,PhysRevD.88.072003,Aaltonen:2014eva} are reduced,
while the latest measurements by the D0 collaboration~\cite{Abazov:2014oea,D02} are now in good
agreement with the predictions.

This paper is organized as follows. In~\secRef{sec:Detector} the main components
of the \atlas{} detector relevant for this measurement are summarized.
In~\secRef{sec:Samples} the simulated samples used for the analysis are
presented. In~\secRef{sec:Selection} the object and event selection are
described. In~\secRef{sec:Reconstruction} the kinematic reconstruction used for
the \Actt measurement is detailed. For comparison with theory prediction, the
measurements are corrected for detector resolution and acceptance effects, as
presented in~\secRef{sec:Corrections}. \SecsAndSecRef{sec:Systematics}{sec:Results} 
describe the systematic uncertainties and the measurement results, respectively. 
Finally, the conclusions are given in~\secRef{sec:Conclusion}.

\section{The ATLAS detector}
\label{sec:Detector}
The ATLAS detector~\cite{atlas} at the LHC covers nearly the entire solid angle 
around the collision point.\footnote{ATLAS uses a right-handed  
coordinate system with its origin at the nominal 
interaction point (IP) in the centre of the detector and the $z$-axis along 
the beam pipe. The $x$-axis points from the IP to the centre of the LHC ring, 
and the $y$-axis points upward. Cylindrical coordinates $(r,\phi)$ are used 
in the transverse plane, $\phi$ being the azimuthal angle around the beam
pipe.}
It consists of an inner tracking detector surrounded 
by a thin superconducting solenoid, electromagnetic and hadronic calorimeters, 
and a muon spectrometer incorporating three large superconducting toroid magnets.
The inner-detector system is immersed in a 2 T axial magnetic field 
and provides charged-particle-tracking in the range $|\eta| < 2.5$. 

A high-granularity silicon pixel detector covers the interaction region and typically 
provides three measurements per track. 
It is surrounded by a silicon microstrip tracker designed to provide four 
two-dimensional measurement points per track. These silicon detectors are complemented 
by a transition radiation tracker, which enables radially extended track reconstruction 
up to $|\eta| = 2.0$. The transition radiation tracker also provides electron 
identification information based on the fraction of hits (typically 30 in total) 
exceeding an energy-deposit threshold corresponding to transition radiation.

The calorimeter system covers the pseudorapidity range $|\eta|< 4.9$. 
Within the region $|\eta|< 3.2$, electromagnetic calorimetry is provided 
by barrel and end-cap high-granularity lead/liquid-argon (LAr) electromagnetic 
calorimeters, with an additional thin LAr presampler covering $|\eta| < 1.8$ 
to correct for energy loss in the material upstream of the calorimeters. Hadronic 
calorimetry is provided by a steel/scintillator-tile calorimeter, segmented 
into three barrel structures within $|\eta| < 1.7$, and two copper/LAr hadronic 
endcap calorimeters. The solid angle coverage is completed with forward copper/LAr 
and tungsten/LAr calorimeters used for electromagnetic and hadronic measurements.

The muon spectrometer comprises separate trigger and high-precision tracking 
chambers measuring the deflection of muons in a magnetic field generated by 
superconducting air-core toroids. The precision chamber system covers the 
region $|\eta| < 2.7$ with three layers of monitored drift tube chambers, complemented 
by cathode strip chambers in the forward region. 
The muon trigger system covers the range $|\eta| < 2.4$ with resistive plate 
chambers in the barrel, and thin gap chambers in the endcap regions.

A three-level trigger system is used to select interesting events. 
The Level-1 trigger is implemented in hardware and uses a subset of detector information 
to reduce the event rate to a design value of at most 75~kHz. This is followed by two 
software-based trigger levels, which together reduce the event rate to about 300~Hz.

\section{Simulated samples}
\label{sec:Samples}
Several Monte Carlo (MC) simulated samples are used in the analysis to model the signal and background processes. 
The total background, estimated partly from these simulated samples, is subtracted from the data at a later stage of the analysis. The signal sample is used to correct the background subtracted data for detector, resolution and acceptance effects. The MC samples are also used to evaluate the systematic uncertainties of the measurement.

The nominal simulated \ttbar{} sample is generated using the
\powheg~\cite{FRI-0701,Nason:2004rx,Frixione:2007nw} (patch4) generator with the
CT10~\cite{Lai:2010vv} parton distribution function (PDF) set. The NLO QCD
matrix element is used for the \ttbar{} hard-scattering process. The parton
showers (PS) and the underlying event are simulated using
\pythia{}~\cite{SJO-0601} (v6.425) with the CTEQ6L1~\cite{cteq6} PDF and the
corresponding Perugia 2011C set of tunable parameters (tune)~\cite{PhysRevD.82.074018} intended to be used with this PDF. The hard-scattering process renormalization and factorization scales are fixed at the generator default value $Q$ that is defined by 
\begin{equation}
Q=\sqrt{\mt^2+\pt^2},
\label{eq:hps_powheg}
\end{equation}
where $\mt$ and $\pt$ are the top quark mass and the top quark transverse momentum, evaluated for the underlying Born configuration (i.e. before radiation).  Additional \ttbar{} samples used to evaluate signal modelling uncertainties are described in \secRef{sec:Systematics}.
Signal samples are normalized to a reference value of $\sigma_{t\bar{t}}= 177^{+10}_{-11}$~pb for a top 
quark mass of $\mt=172.5~\GeV$. The cross-section has been calculated at next-to-next-to-leading-order (NNLO) in QCD including resummation of next-to-next-to-leading logarithmic 
(NNLL) soft gluon terms~\cite{Cacciari:2011hy, Baernreuther:2012ws, Czakon:2012zr,Czakon:2012pz, Czakon:2013goa, Beneke:2011mq} with top++2.0~\cite{Czakon:2011xx}. The PDF and strong coupling ($\alpha_{\rm{s}}$) uncertainties 
were calculated using the PDF4LHC prescription~\cite{Botje:2011sn} with the MSTW2008 68\% CL 
NNLO~\cite{Martin:2009iq, Martin:2009bu}, CT10 NNLO~\cite{Lai:2010vv, Gao:2013xoa} and 
NNPDF2.3 5f FFN~\cite{Ball:2012cx} PDF sets, and added in quadrature 
to the scale uncertainty. The NNLO+NNLL cross-section value is about 3\% larger than the exact 
NNLO prediction, as implemented in Hathor 1.5~\cite{Aliev:2010zk}.

The MC generators which are utilized to estimate the backgrounds are as follows.  
Single-top processes in the $Wt$ channel are generated with the \mcatnlo{} event
generator (v4.01)~\cite{FRI-0201,Frixione:2008yi} with the CT10 PDF. The parton
showers, hadronization and the underlying event are modelled using the \herwig{}
(v6.520)~\cite{Marchesini:1991ch,COR-0001} and \jimmy{} (v4.31)~\cite{JButterworth:1996zw}
generators. The CT10 PDF with the corresponding \atlas{} AUET2
tune~\cite{PUB-2011-008} is used for parton shower and hadronization settings.
For $Z/\gamma^*$+jets and diboson events
($WW$, $WZ$ and $ZZ$), \alpgen (v2.13)~\cite{MAN-0301} interfaced to \herwig
and \jimmy is used. The CTEQ6L1 PDF and the corresponding \atlas AUET2 tune is
used for the matrix element and parton shower settings. The $Wt$ background
process is normalized to the reference NLO+NNLL QCD~\cite{Kidonakis:2010ux}
prediction. Diboson production is normalized to the reference NLO QCD prediction
obtained using \mcfm~\cite{Campbell:1999ah} and \mcatnlo{} generators with the
MSTW2008 NLO PDF~\cite{Martin:2009iq}. The $Z/ \gamma^* \to ee/\mu\mu$+jets 
cross-section is normalized using a control region in data as detailed
in~\secRef{sec:Selection}. The $Z/ \gamma^* \to \tau\tau$+jets events are
normalized to a NNLO reference cross-section using the
\fewz~\cite{Anastasiou:2003ds} and \zwprod~\cite{Hamberg:1990np} programs with
the MSTW2008~NNLO~PDF.

To realistically model the data, the simulated samples are generated with an average of
eight additional inelastic $pp$ interactions from the same bunch crossing
(referred to as pileup) overlaid on the hard-scatter event. Simulated samples
are processed through ATLAS detector simulation. For the majority of the samples,
a full detector simulation~\cite{atlasfullsim} based on
GEANT4~\cite{Agostinelli:2002hh} is used. Some of the samples used for
assessment of generator modelling uncertainties are obtained using a faster
detector simulation program that relies on parameterized showers in the calorimeters~\cite{atlasfullsim,atlasfastsim}. Simulated events are then processed using the same reconstruction algorithms and analysis chain as the data.

\section{Object and event selection}
\label{sec:Selection}
The data sample collected by the ATLAS detector in 2011 at a 
centre-of-mass energy of 7\,\TeV\ is used for the analysis. The integrated luminosity of the sample is \lumi{} with an overall uncertainty of 1.8\%~\cite{Aad:2013ucp}. The analysis makes use of reconstructed electrons, muons, jets and missing transverse momentum in the detector. 
Electrons are reconstructed as clusters of energy deposits in the electromagnetic calorimeter, matched to a 
track in the inner detector. They are required to pass a set of tight selection criteria~\cite{Aad:2014fxa}.
The selected electrons have to satisfy a requirement on their transverse energy (\et) and the
pseudorapidity of the associated calorimeter cluster ($|\eta_{\rm{cluster}}|$):~$\et> 25$\,\GeV\ and $|\eta_{\rm{cluster}}|<2.47$. 
The electrons in the region $1.37< |\eta_{\rm{cluster}}|<1.52$, which corresponds to a transition 
between the barrel and endcap electromagnetic calorimeters, are excluded. 
Electrons are required to be isolated, using the requirements described as follows (excluding calorimeter 
deposits and tracks from the electrons). The \et{} within a cone of size $\Delta R = \sqrt{(\Delta\eta)^2 + (\Delta\phi)^2} = 0.2$ and the scalar 
sum of track $\pt$ within a cone of $\Delta R=0.3$ around the electron are required to be below $\et$- and 
$\eta$-dependent thresholds. The efficiency of this isolation requirement on electrons is 90\%, and its goal is to reduce the contribution 
from hadrons mimicking lepton signatures, as well as leptons produced in heavy-hadron decays 
or photon conversion. These are referred to as fake and non-prompt leptons (NP) in the following.

Muons are reconstructed by matching a track in the inner detector to a track segment in the muon 
spectrometer. They are required to pass tight selections~\cite{Aad:2014zya}. The selected muons are 
required to have $\pt>20$\,\GeV\ and $|\eta|<2.5$. To reject fake and non-prompt muons, the following isolation 
requirements are imposed:  the calorimeter transverse energy within a cone of $\Delta R = 0.2$ around the muon is required to be less than  $4$\,\GeV\ and the scalar sum of track $\pt$ within a cone of $\Delta R = 0.3$ is required to be less than $2.5$\,\GeV\ (excluding the calorimeter deposits and tracks from the muons).

Jets are reconstructed from energy deposits in the calorimeter, using the anti-$k_t$ algorithm with a distance parameter $R=0.4$~\cite{Cacciari:2008gp}.
The energy of the  input clusters~\cite{atlastopo} is corrected to the level of
stable particles using calibration factors derived from simulation and data~\cite{Aad:2014bia}.
The jets are required to have a \pt{} of at least $25$\,\GeV\ and $|\eta|<2.5$. To suppress the contribution from low-$\pt$ jets originating from pileup interactions, tracks associated with the jet and emerging from the primary vertex are required to account for at least 75\% of the scalar sum of the $\pt$ of all tracks associated with the jet. A primary vertex, originating from $pp$ interactions, is a reconstructed vertex required to have at least five associated tracks with $\pt>0.4$\,\GeV.
In the cases where more than one primary vertex is reconstructed, the vertex with the highest $\sum_{\rm trk} \pt^2$ is chosen and assumed to be associated with the hard-process, and the sum runs over all associated tracks.

The missing transverse momentum (\met) is a measure of transverse momentum imbalance due to the presence of neutrinos. It is reconstructed from the transverse momenta of jets in the kinematic range of $\pt>20$\,\GeV\ and $|\eta|<4.5$, electrons, muons, and calorimeter clusters not associated with any of the reconstructed objects, as detailed in \citeRef{Aad:2012re}.

Using the objects reconstructed as above, an event selection optimized for signatures corresponding to \ttbar{} events in which both $W$ bosons from the $t$ and $\tbar$ quarks decay to leptons is performed. 
Events are required to have been selected by a single-electron trigger with a threshold of $20$ or $22$\,\GeV\ (depending on
the data-taking period), or a single-muon trigger with a threshold of $18$\,\GeV. They are required to have exactly two isolated, oppositely charged, leptons. Depending on the lepton flavours, the sample is divided into three analysis channels referred to as \ee, \emu{} and \mumu{}. To reduce the Drell--Yan production of $Z/ \gamma^*$+jets background, the invariant mass of the two leptons ($m_{\ell\ell}$) is required to be above a threshold used to suppress $\gamma^* \to \ell\ell$ production background and outside a $Z$ boson mass window in the \ee and \mumu{} channel events. The following requirements are used: $m_{\ell\ell} > 15$\,\GeV\ and $|m_{\ell\ell} - m_Z| > 10$\,\GeV. In the \ee and \mumu{} channels the Drell--Yan and diboson backgrounds are further reduced using a requirement on the missing transverse momentum, $\met > 60$\,\GeV. In the \emu{} channel the $Z/ \gamma^*$+jets background is smaller and suppressed by requiring the scalar sum of the \pt{} of the two leading jets and leptons ($\HT{}$) to be larger than 130\,\GeV.

The background contributions are estimated using a combination of techniques using data and Monte Carlo events.
In the case of single-top and diboson processes, both the shape and normalization of the distributions are taken from the simulation. For $Z/ \gamma^* \to ee/\mu\mu$+jets events, 
simulated MC events are used to model the shape of the distributions, but a data control
region is used for normalization. 
Drell--Yan events with $\met >  60$\,\GeV\ are affected by energy mismeasurements, that are difficult to model in simulation.
A control region with events with $m_{\ell\ell}$ in the $Z$-mass region is defined to study the effect of mismeasured \met.
The relative \met, defined as the projection of the missing transverse momentum onto the direction of the jet or charged lepton with closest $\phi$, is used to identify the events with mismeasured objects. 
Events with energy mismeasurements are characterized by high values of relative \met.
A cut is applied to the relative \met, and data and simulation are then compared to derive a normalization correction
factor which is applied to the simulated sample. The $Z/ \gamma^* \to \tau\tau$ contribution is estimated from MC simulation. 
The background stemming from events with at least one non-prompt or fake lepton is estimated from the data, 
since the lepton misidentification rates are difficult to model in MC simulation.
A matrix method technique is used~\cite{Aad:2010ey}. It consists of selecting data samples dominated either by real leptons or by fake leptons, and estimating the efficiencies for a real or fake lepton to satisfy the isolation criteria.

\begin{figure}[htp]
\centering
\subfigure[]{
        \includegraphics[width=0.47\textwidth]{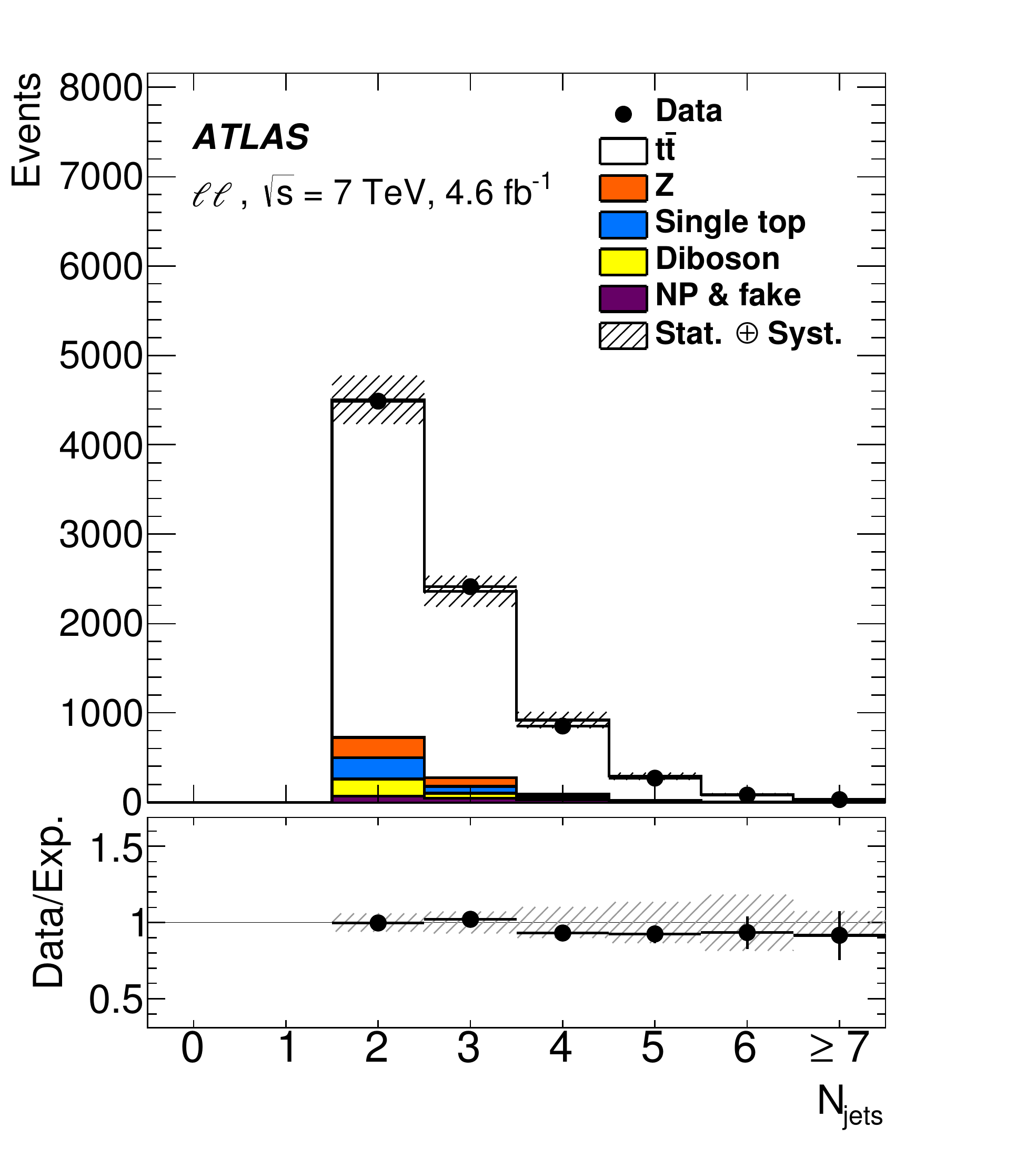}
	\label{fig:dataMC:njt}
} 
\subfigure[]{
 	\includegraphics[width=0.47\textwidth]{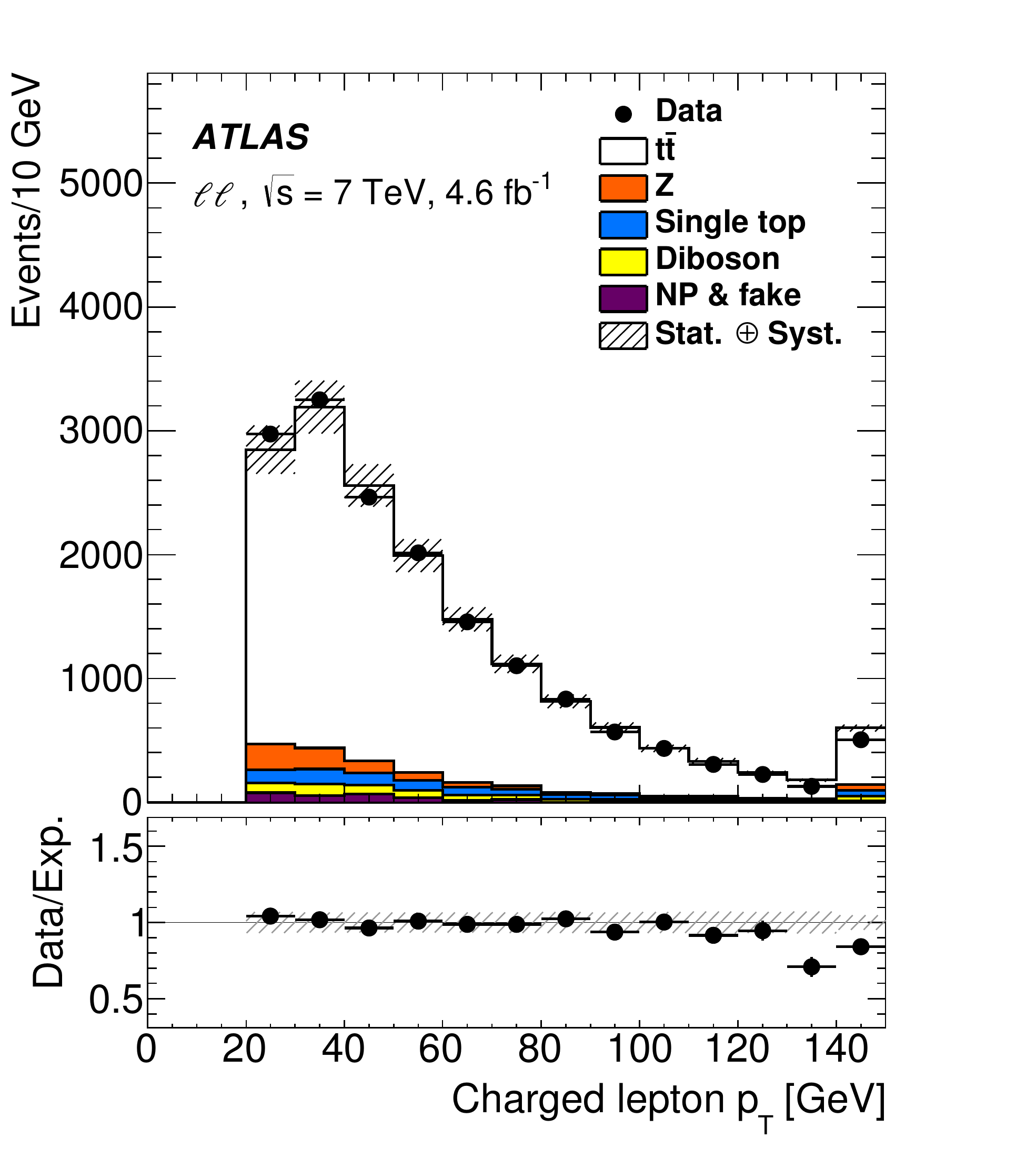}
	\label{fig:dataMC:leppt}
}
\subfigure[]{
 	\includegraphics[width=0.47\textwidth]{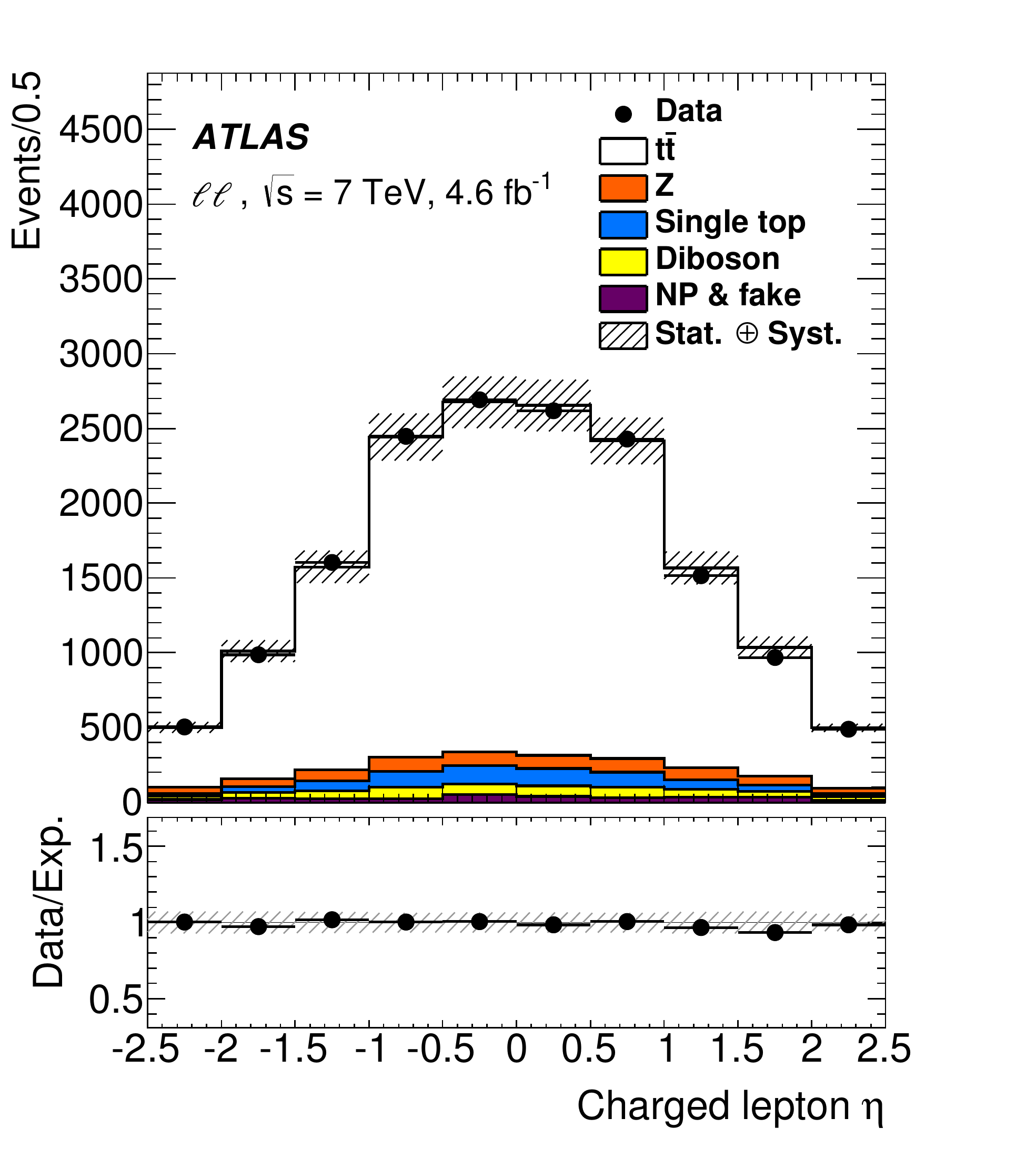}
	\label{fig:dataMC:lepeta}
}
\subfigure[]{
 	\includegraphics[width=0.47\textwidth]{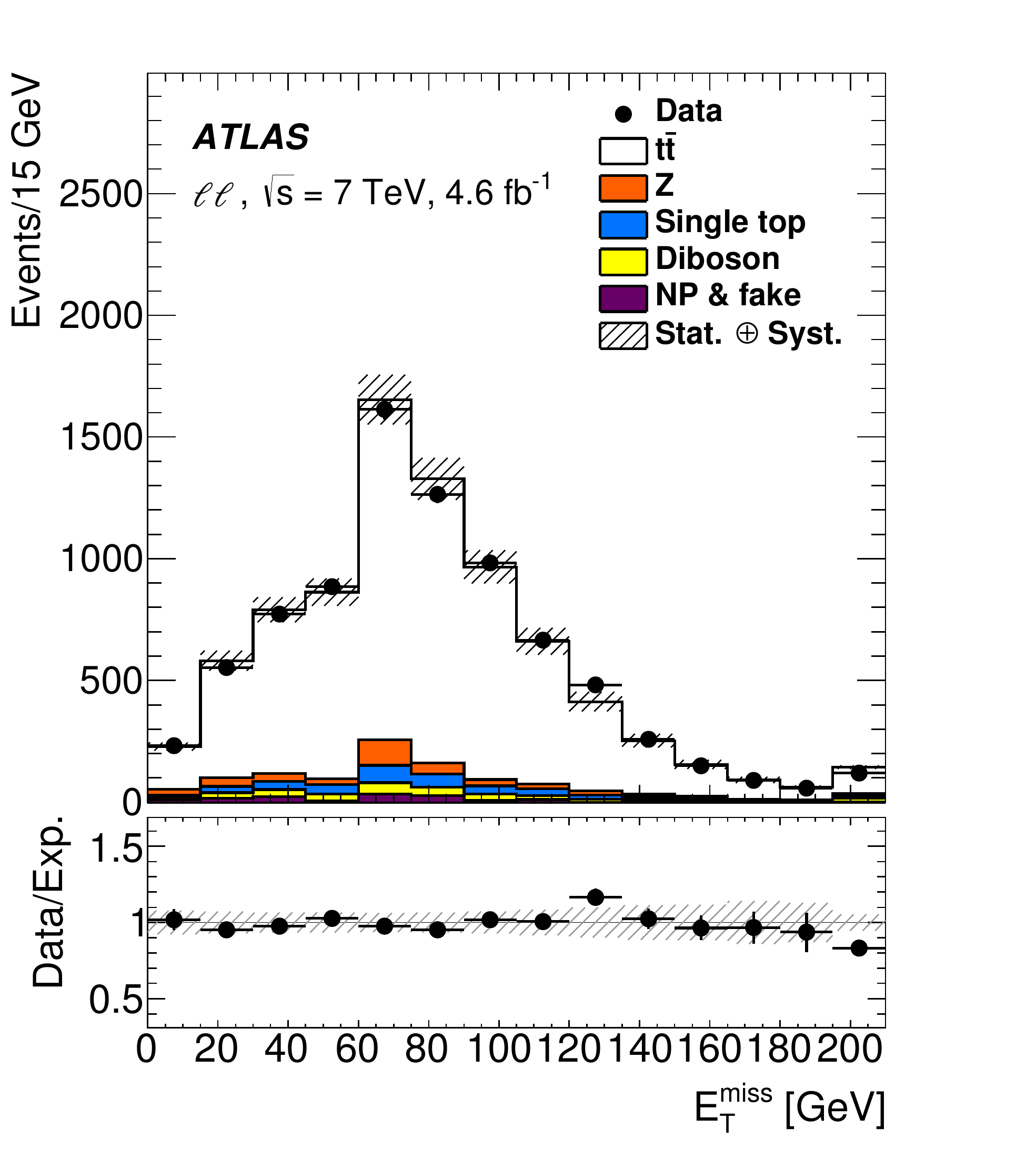}
	\label{fig:dataMC:met}
}
\caption{\label{fig:dataMC} Comparison of the expected and observed distributions of (a)~the jet multiplicity, 
         (b)~the lepton transverse momentum \pt{}, (c)~the lepton pseudorapidity $\eta$ and (d)~the missing transverse momentum \met, shown 
         for the combined \ee, \emu and \mumu channels. Events beyond the range of the horizontal axis of (a), (b) 
         and (d) are included in the last bin. The hatched area corresponds to the combined statistical and systematic uncertainties.
         Events with one or more non-prompt or fake leptons are referred to as "NP \& fake".}
\end{figure}

\begin{figure}[htp]
\centering
\subfigure[] {
        \includegraphics[width=.47\textwidth]{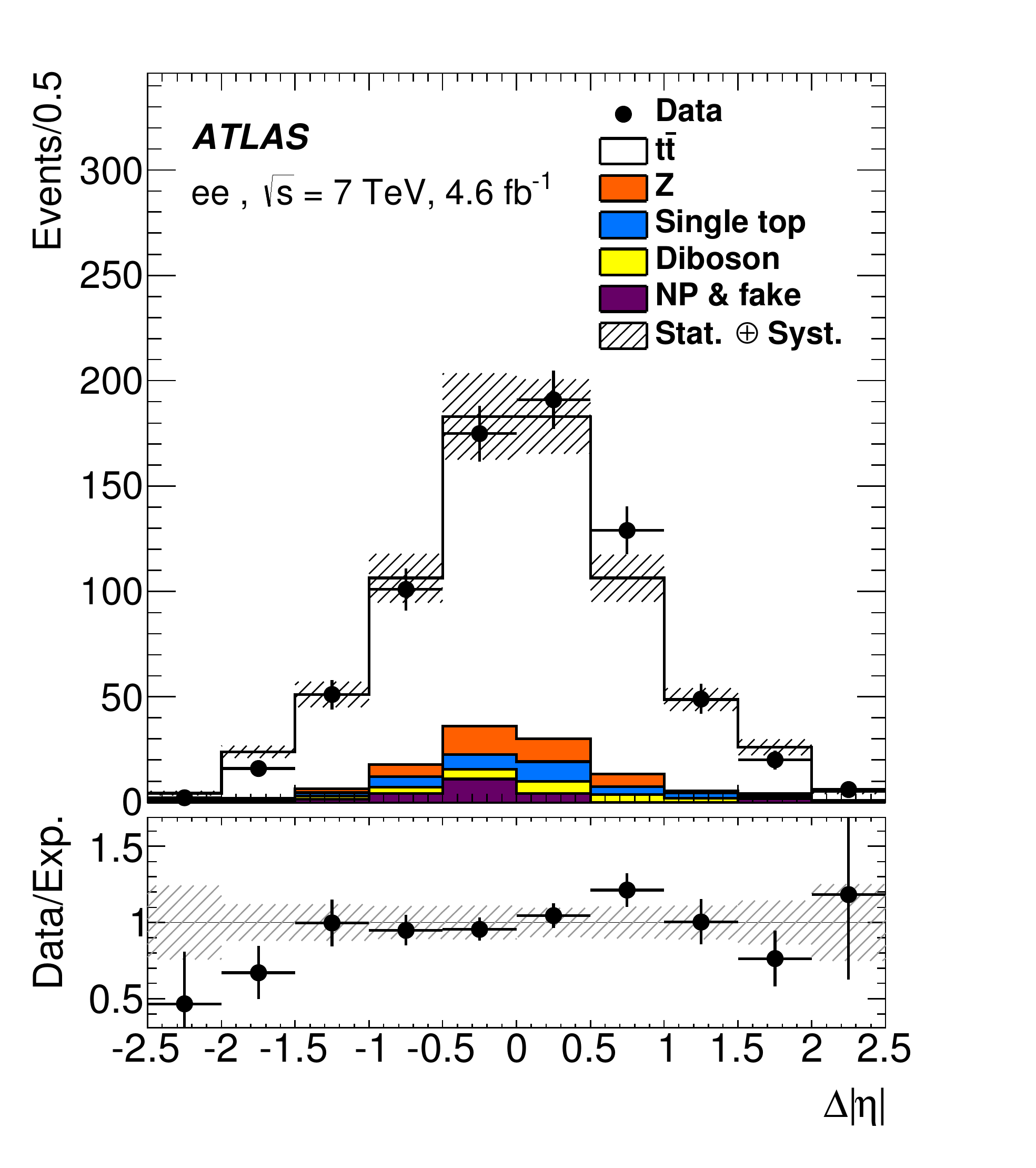}
\label{fig:dataMC_deta:ee}
}
\subfigure[] {
\includegraphics[width=.47\textwidth]{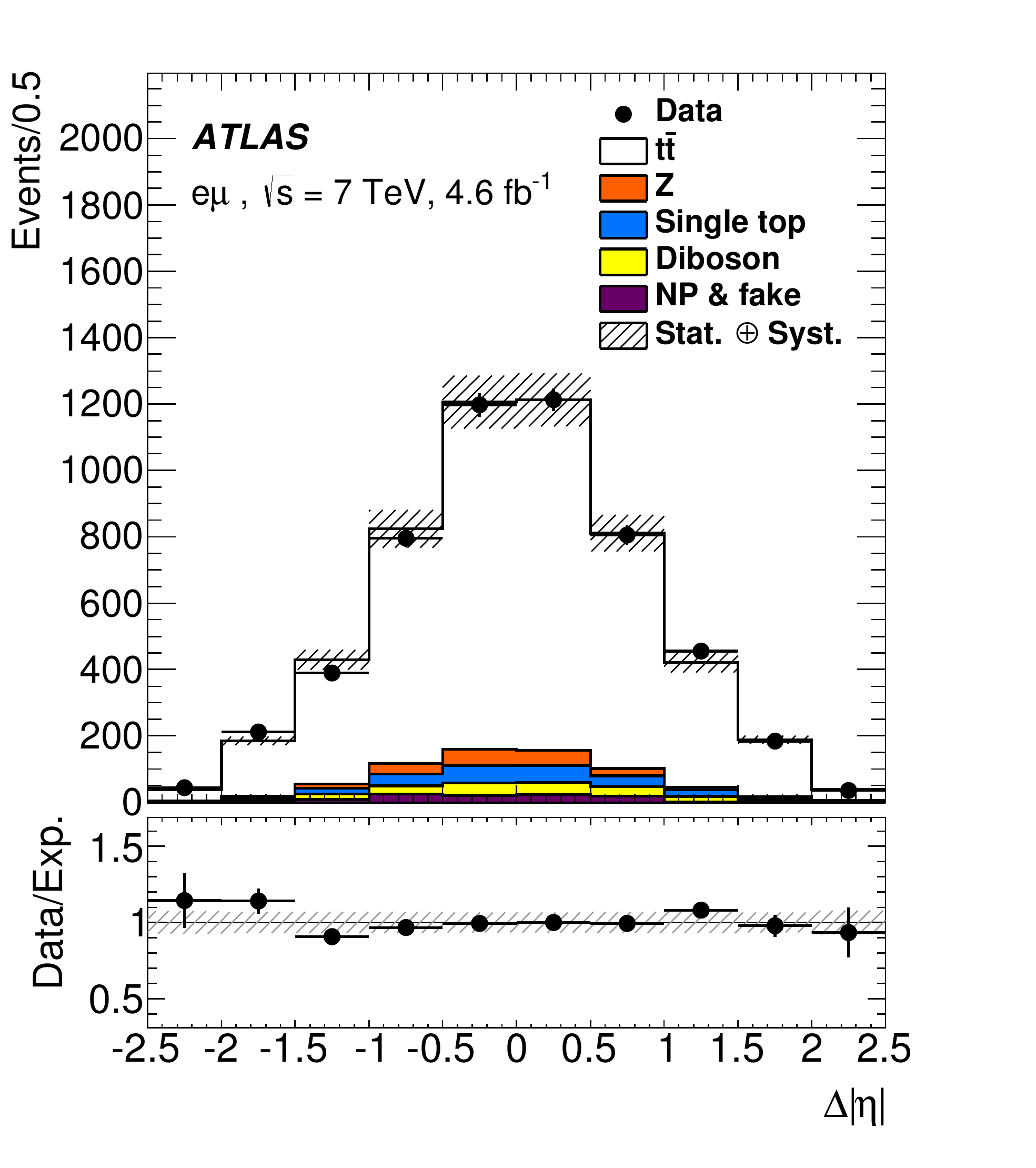}
	\label{fig:dataMC_deta:emu}
}
\subfigure[] {
\includegraphics[width=.47\textwidth]{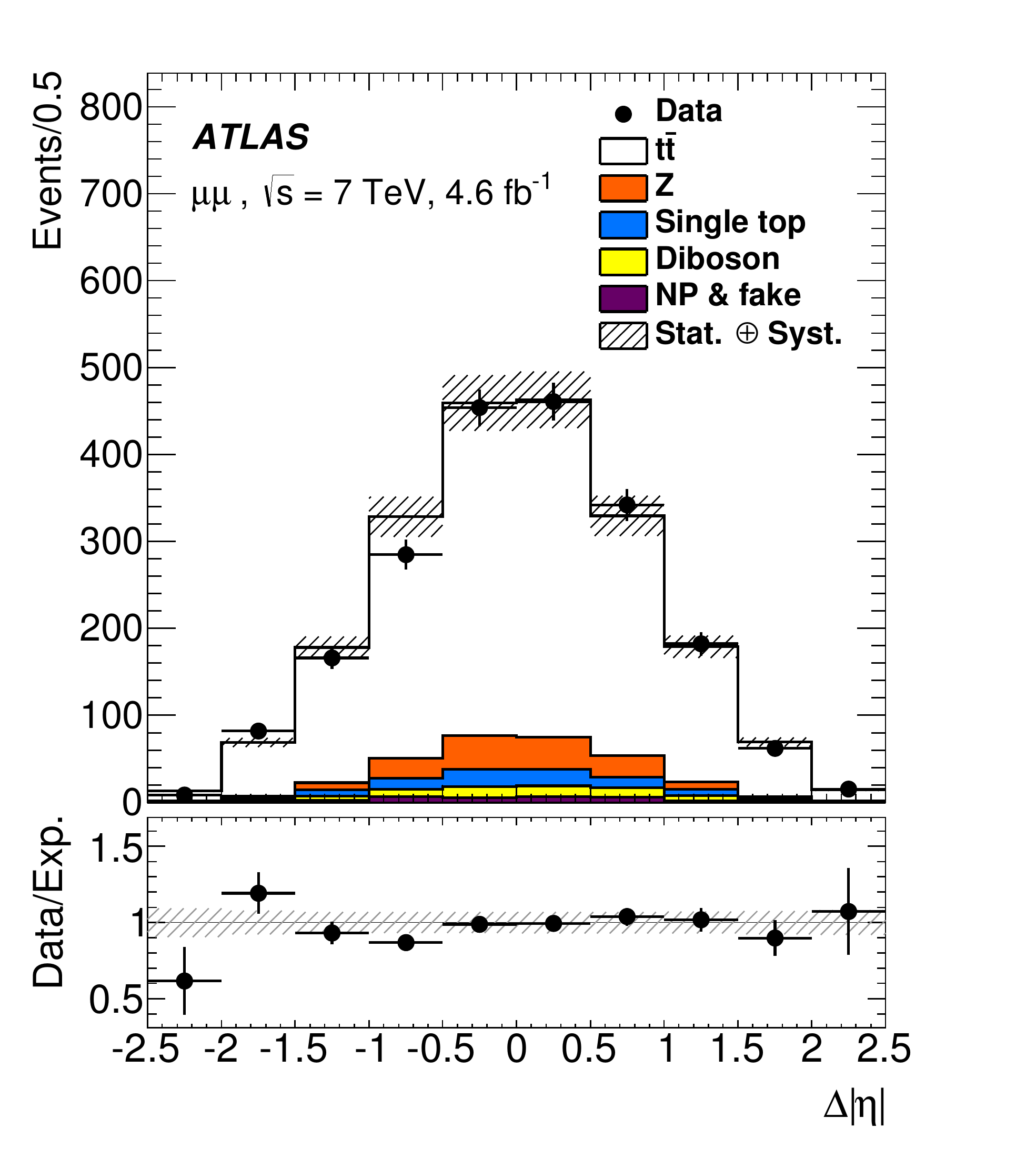}
	\label{fig:dataMC_deta:mumu}
}
\caption{\label{fig:dataMC_deta} Comparison of the expected and observed distributions of the \deta{} 
         variable for the (a) \ee, (b) \emu and (c) \mumu channels. The hatched area corresponds to the 
         combined statistical and systematic uncertainties.
         Events with one or more non-prompt or fake leptons are referred to as "NP \& fake".}
\end{figure}

After the final selection, the data sample contains more than 8000 events, with an expected 
signal-to-background ratio of approximately six. The number of events in data and simulation, 
including statistical and systematic uncertainties, are compared in \tabRef{tab:evtnumbers}.
After selection, the largest number of events is observed in the \emu{} channel, which has the 
highest branching ratio and the loosest background suppression cuts.
The \ee{} channel has the lowest number of events because of the stringent 
requirements on lepton kinematics. 
Figure~\ref{fig:dataMC} shows good agreement between the data and the SM predictions  
for the jet multiplicity, lepton \pt{} and lepton pseudorapidity distributions.
The \deta{} distributions are shown in \figRef{fig:dataMC_deta} for the three channels separately. 

\begin{table}[tbp]
\centering
    \begin{tabular}{| l | rll | rll | rll  |}
    \hline
    Channel                 & \multicolumn{3}{c|}{$ee$}            & \multicolumn{3}{c|}{$e\mu$}         & \multicolumn{3}{c|}{$\mu\mu$}  \\
    \hline
    \ttbar{}                & 621  & $\pm$ 5   & $\pm$ 59     & 4670 & $\pm$ 10 & $\pm$ 325   & 1780 & $\pm$ 10  & $\pm$ 120   \\
    Single top              & 31.6 & $\pm$ 1.7 & $\pm$ 3.8    & 230  & $\pm$ 5  & $\pm$ 21    & 83.9 & $\pm$ 2.7 & $\pm$ 8.3   \\
    Diboson                 & 22.8 & $\pm$ 0.9 & $\pm$ 2.6    & 177  & $\pm$ 3  & $\pm$ 16    & 61.5 & $\pm$ 1.5 & $\pm$ 6.1   \\
    \Zee\ (DD)              & 20.8 & $\pm$ 1.7 & $\pm$ 1.4    & \multicolumn{3}{c|}{\allzero} & \multicolumn{3}{c|}{\allzero}  \\
    \Zmm\ (DD)              & \multicolumn{3}{c|}{\allzero}   & 2.1  & $\pm$ 0.5 & $\pm$ 0.7  & 77   & $\pm$ 4   & $\pm$ 12    \\
    \Ztau{}                 & 18.6 & $\pm$ 1.8 & $\pm$ 7.0    & 170  & $\pm$ 6   & $\pm$ 60   & 67   & $\pm$ 4   & $\pm$ 25    \\
    NP \& fake (DD)         & 19   & $\pm$ 4   & $\pm$ 19     & 99   & $\pm$ 10 & $\pm$ 63    & 26.8 & $\pm$ 5.1 & $\pm$ 1.9   \\
    \hline
    Total expected          & 734  & $\pm$ 8   & $\pm$ 63     & 5350  & $\pm$ 20  & $\pm 340$           & 2100 & $\pm$ 10  & $\pm 130$ \\
    \hline
    Data                    & \multicolumn{3}{c|}{740}            & \multicolumn{3}{c|}{5328}            & \multicolumn{3}{c|}{2057} \\
    \hline
    \end{tabular}

\caption{\label{tab:evtnumbers} Observed number of data events in comparison to the expected number
  of signal events and all relevant background contributions after the event selection. The backgrounds are estimated from the MC simulation or from the data-driven methods~(DD) described in \secRef{sec:Selection}.
  Events with one or more non-prompt or fake leptons are referred to as "NP \& fake".
  The first uncertainty is statistical, the second corresponds to systematic
  uncertainties on background normalization and detector modelling described in \secRef{sec:Systematics}. The values labeled with "\allzero{}" are estimated to be smaller than 0.5.}
\end{table}

\section{Kinematic reconstruction}
\label{sec:Reconstruction}
For the measurement of the $\ttbar$ charge asymmetry, the direction of the top and antitop quarks
needs to be determined. The four-momenta of top quarks in selected events are computed
with a kinematic reconstruction using the objects observed in the detector.
The reconstruction is based on solving the kinematic equations obtained when imposing energy--momentum 
conservation at each of the decay vertices of the process.
In the dilepton channel, at least two neutrinos are produced and escape undetected. 
Consequently, the system is underconstrained and its kinematics cannot be fully determined without further assumptions (for example on the $W$ boson and top quark masses, and the pseudorapidities of the neutrinos from the $W$ boson decays).
Moreover, several ambiguities have to be resolved
to find the correct solution. For example, the lepton and jet from the same decay chain 
have to be associated.
In an event with
two leptons and two jets, this leads to two possible associations.
In this analysis, the neutrino weighting
technique~\cite{Abbott:1997fv} is used. 
This procedure steps through different hypotheses for the pseudorapidity
of the two neutrinos in the final state. 
These hypotheses are made independently for the two neutrinos.
For each
hypothesis, the algorithm calculates the full event kinematics, assuming the $W$ boson and the top quark masses.
It then assigns a weight to the resulting solution based
on the level of agreement between the calculated and measured missing
transverse momentum. 
The weight is defined as
\begin{equation}
 w = \prod_{d=x,y} \exp\left( \frac{-(E_d^{\rm{miss,calc}} - E_d^{\rm{miss,obs}})^2 }{ 2(\sigma_{\et^{\rm{miss}}})^2 }\right),
\end{equation}
with $E_d^{\rm{miss,obs}}$ being the projection of the measured missing transverse momentum along the axes  
defining the transverse plane ($d=x,y$) and $E_d^{\rm{miss,calc}}$ the projection calculated with 
the assumed $\eta$ values of the neutrino pair. The resolution on the missing transverse momentum is 
denoted $\sigma_{\et^{\rm{miss}}}$, and defined as 
$\sigma_{\et^{\rm{miss}}} = 0.5\sqrt{\sum \et} \GeV$~\cite{Aad:2012re}. 
The total transverse energy, $\sum \et$ is defined as $\sum \et = \sum_{i=1}^{N_{\rm{cell}}}E_i \sin{\theta_i}$ 
where $E_i$ and $\theta_i$ are the energy and the polar angle of calorimeter cells associated with clusters.
All possible lepton--jet associations are considered and jet energy mismeasurements are accounted for 
by random shifts of the jet energies within their resolutions. 
The solution corresponding to the maximum weight is
selected to represent the event.

As a result of the scan over neutrino pseudorapidities and jet energy values, the reconstruction efficiency, 
corresponding to the fraction of events in which solutions for $t$ and $\tbar$ four-momenta are found, 
is estimated to be about 80\%~in the data. 
In the other 20\% of events, no solution to the system of the kinematic equations could be found, and the events are not used for the measurement of \Actt.
The performance of the reconstruction algorithm for key variables,
such as the top quark rapidities and $\Delta |y|$,    
is evaluated using the nominal $\ttbar$ simulated sample.
The fraction of reconstructed MC events where the sign of $\Delta|y|$ is determined correctly is about 70\%. 
In the simulated samples, the correct combination of the charged leptons and two jets from $b$($\bar{b}$)-quarks is 
found in approximately 80\% of the events with exactly two reconstructed jets, both of which are matched to the $b$($\bar{b}$)-quarks. 
In case all events passing the event selection are considered, the correct combination is found in approximately 47\% of the events.

\begin{figure}[tbp]
\centering
\subfigure[] {
\includegraphics[width=.47\textwidth]{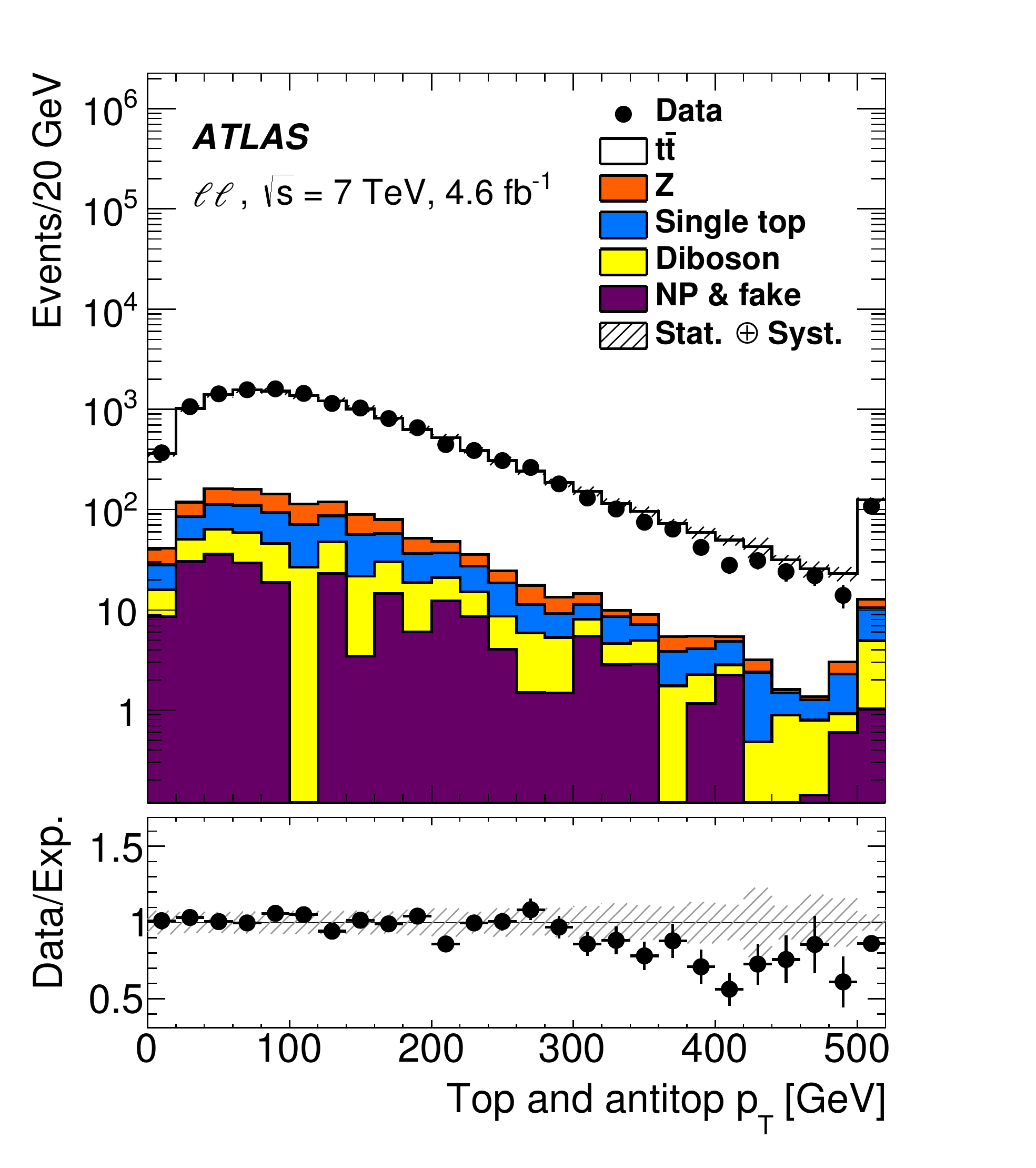}
\label{fig:reco_dataMC:pt}
}       
\subfigure[] {
\includegraphics[width=.47\textwidth]{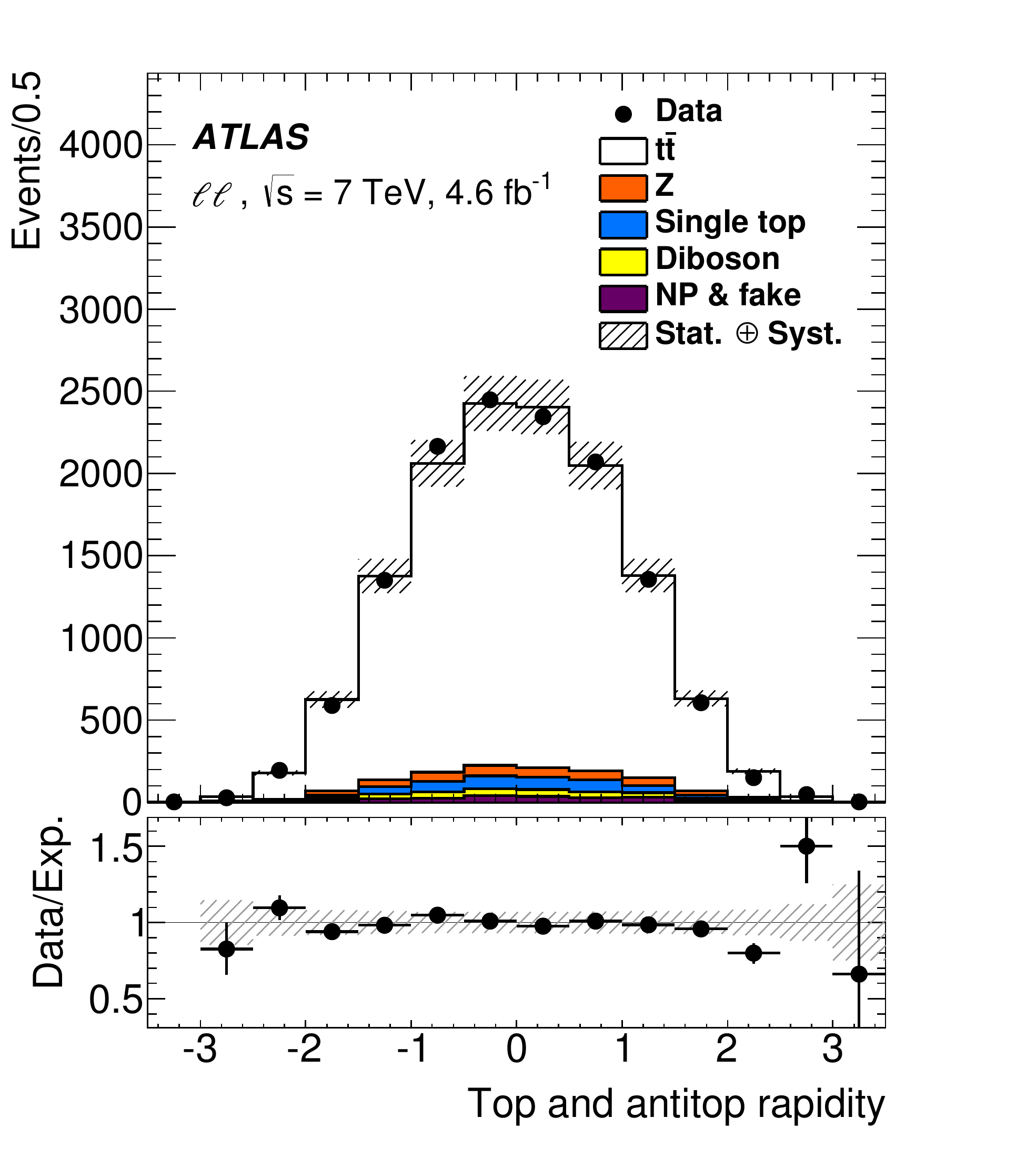}
\label{fig:reco_dataMC:rap}
} 
\subfigure[] {
\includegraphics[width=.47\textwidth]{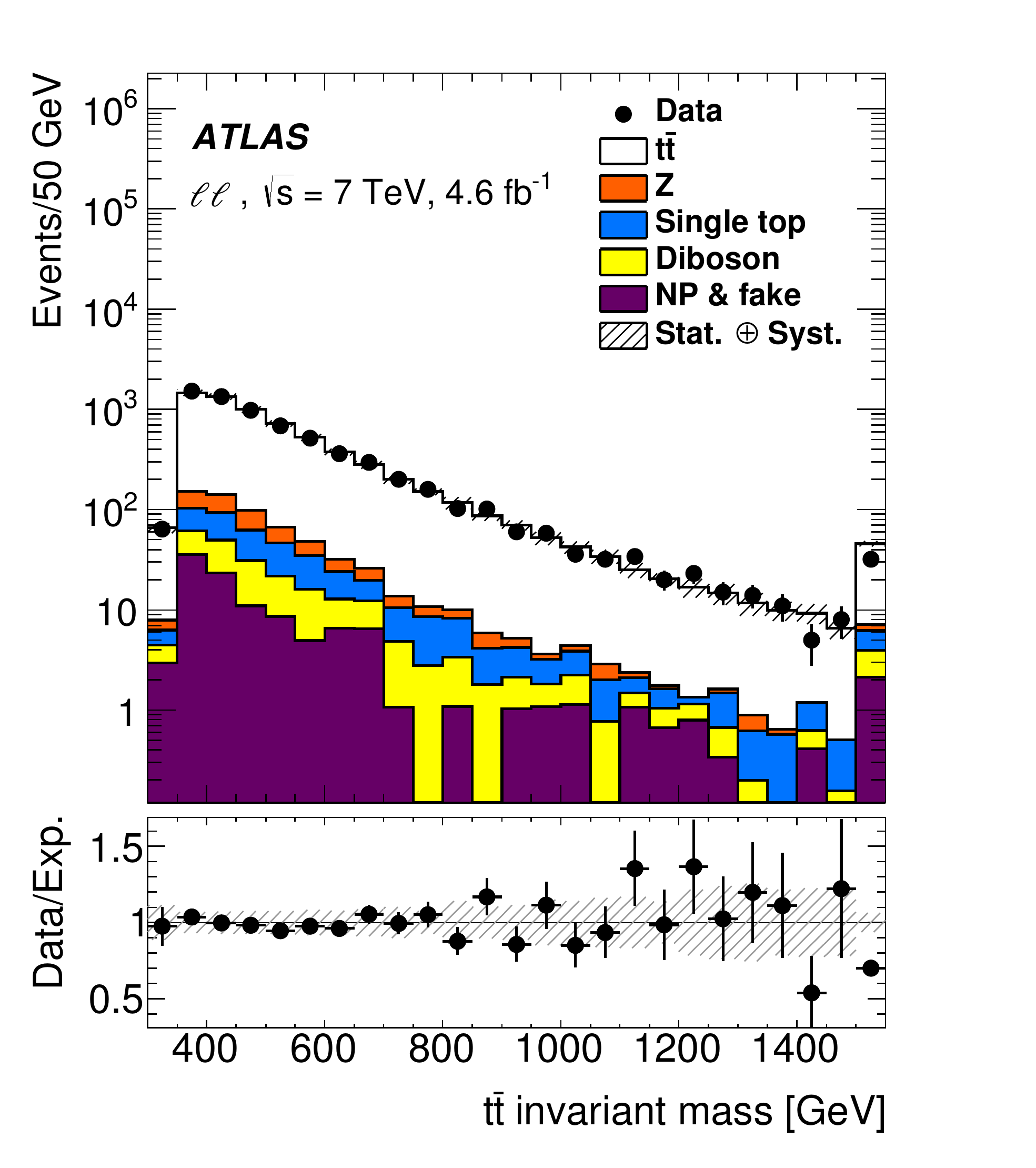}
\label{fig:reco_dataMC:mttbar}
} 
\caption{\label{fig:reco_dataMC} Comparison of the expected and observed distributions of (a)~the top and antitop quark transverse momentum \pt, (b)~top and antitop quark rapidity and (c)~the \ttbar{} invariant mass, shown for the combined \ee,~\emu and \mumu channels. The hatched area corresponds to the combined statistical and systematic uncertainties. Events with one or more non-prompt or fake leptons are referred to as "NP \& fake".}
\end{figure}

\begin{figure}[tbp]
\centering
\subfigure[] {
\includegraphics[width=.47\textwidth]{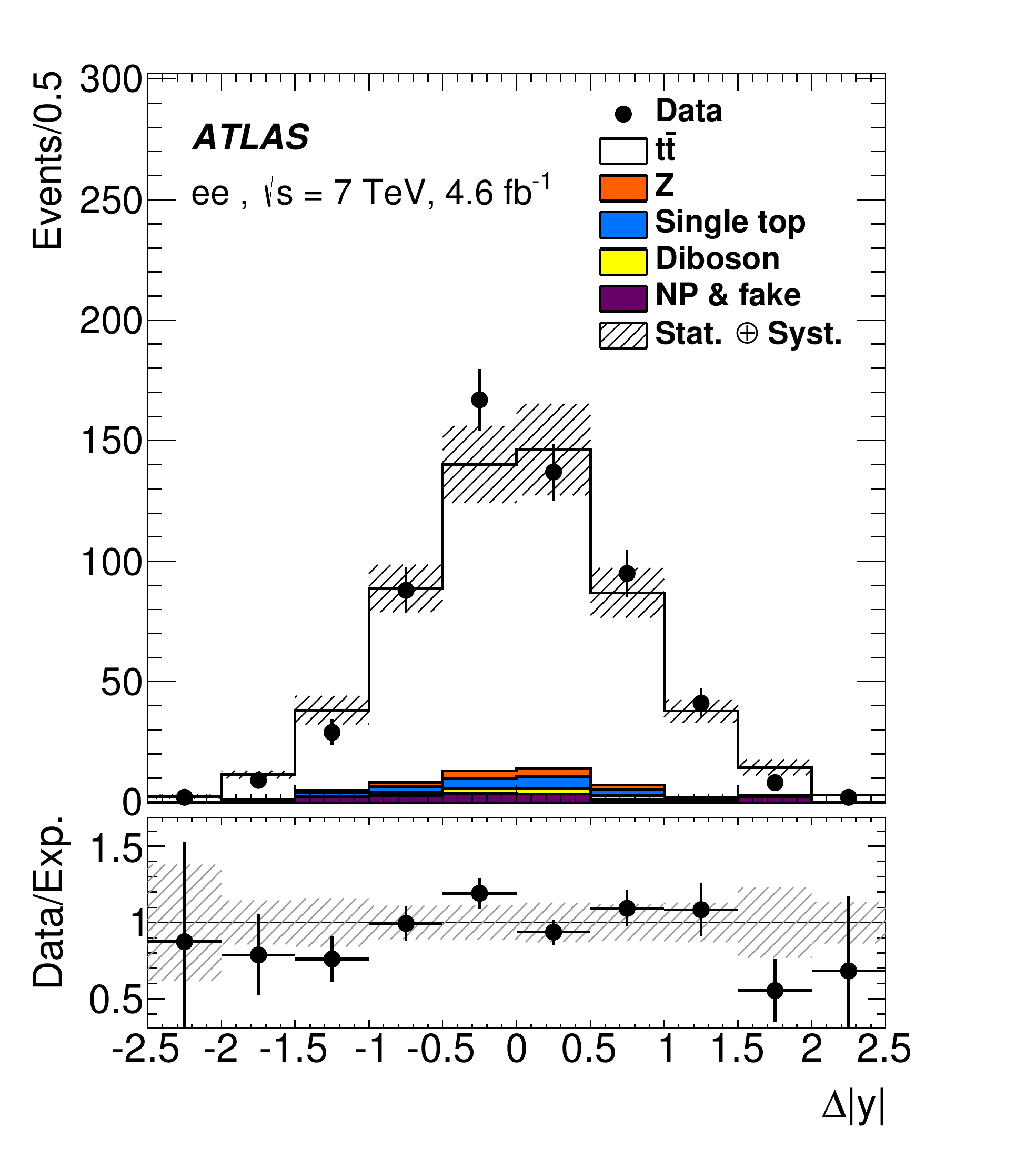}
\label{fig:reco_deltarap:ee}
} 
\subfigure[] {
\includegraphics[width=.47\textwidth]{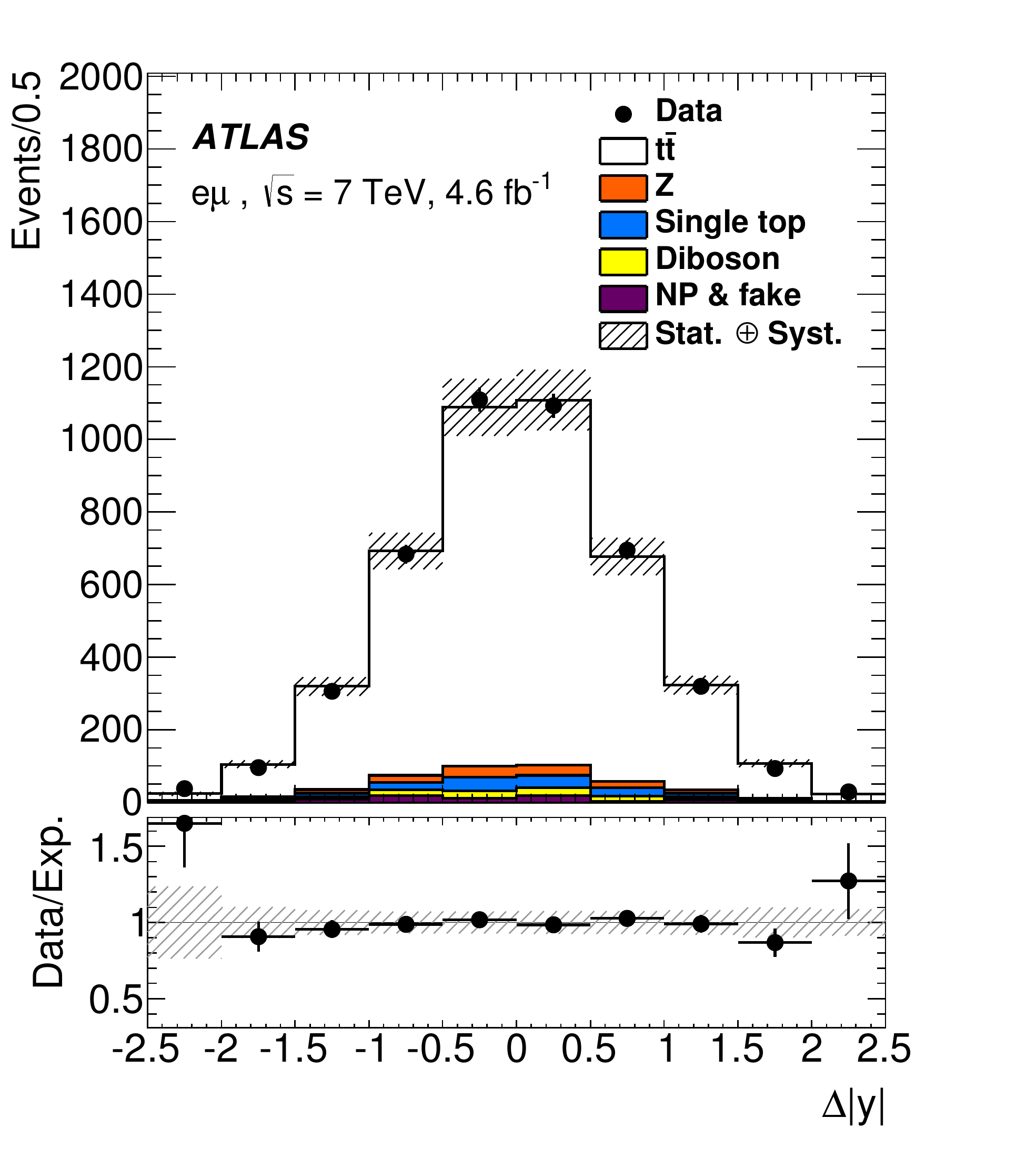}
\label{fig:reco_deltarap:emu}
} 
\subfigure[] {
\includegraphics[width=.47\textwidth]{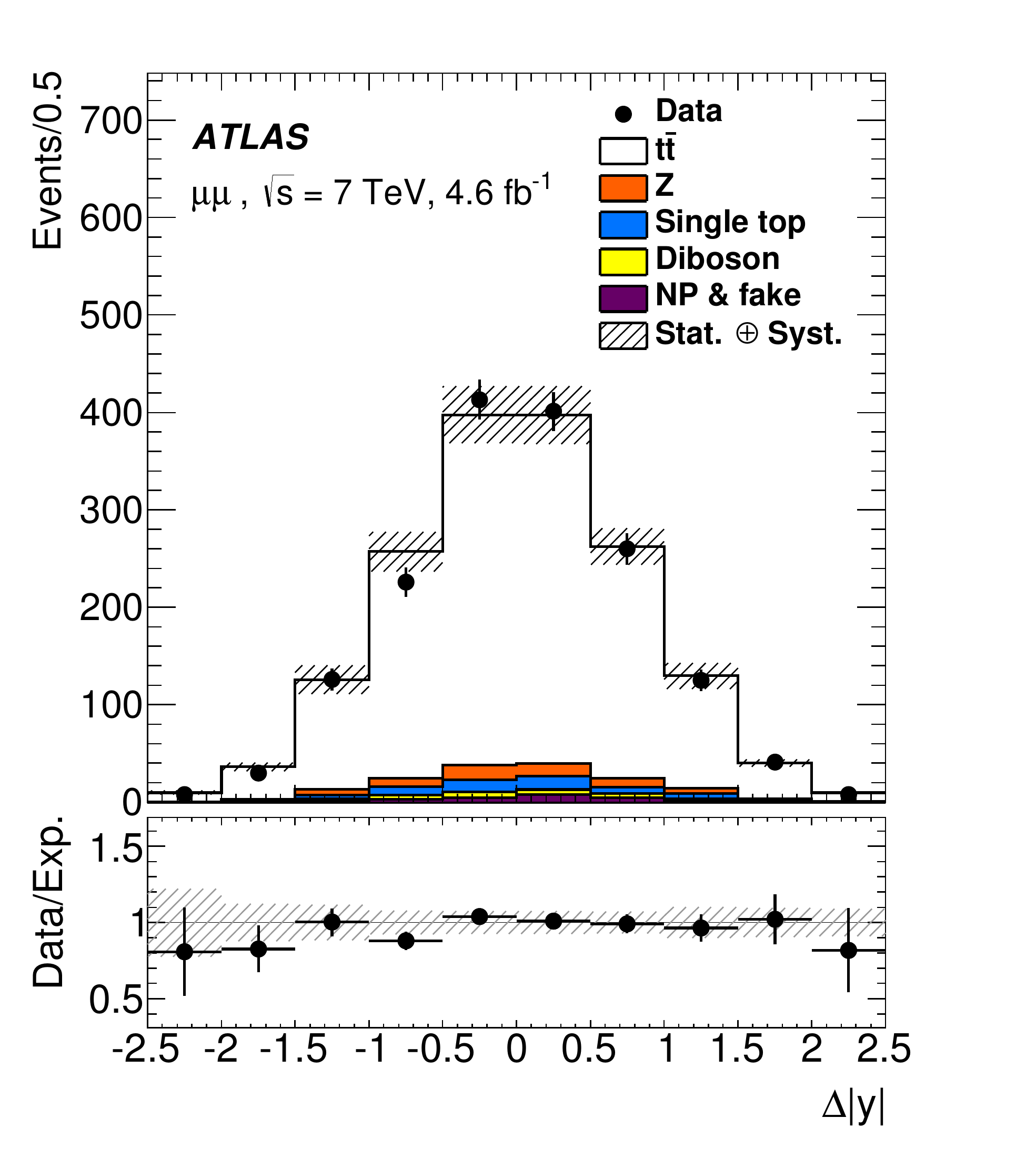}
\label{fig:reco_deltarap:mumu}
} 
\caption{\label{fig:reco_deltarap} Comparison of the expected and observed distributions of the $\Delta|y|$ variable for the (a) \ee, (b) \emu and (c) \mumu channels. The hatched area corresponds to the combined statistical and systematic uncertainties. Events with one or more non-prompt or fake leptons are referred to as "NP \& fake".}
\end{figure}

In \figRef{fig:reco_dataMC} the distributions of the top quark transverse momentum, top quark rapidity and the \ttbar{} invariant mass are shown for the combined \ee,~\emu and \mumu channels. In \figRef{fig:reco_deltarap} the $\Delta|y|$ distribution is shown separately for each of the \ee,~\emu and \mumu channels. Good agreement between the observed and expected distributions is found. 

\clearpage

\section{Corrections}
\label{sec:Corrections}
For comparison with theoretical calculations, the measurements are corrected for detector resolution and acceptance effects.
The corrections are applied to the observed \deta{} and \dy{} spectra. 
Apart from the corrected inclusive asymmetry values, particle- or parton-level 
\deta{} and \dy{} distributions are obtained and presented as normalized 
differential cross-sections in \secRef{sec:Results}. Acceptance corrections are included, 
thus all the results correspond to an extrapolation to the full phase-space for \ttbar{} production. 

\begin{figure}[tbp]
\centering 
\subfigure[] {
\includegraphics[width=.47\textwidth]{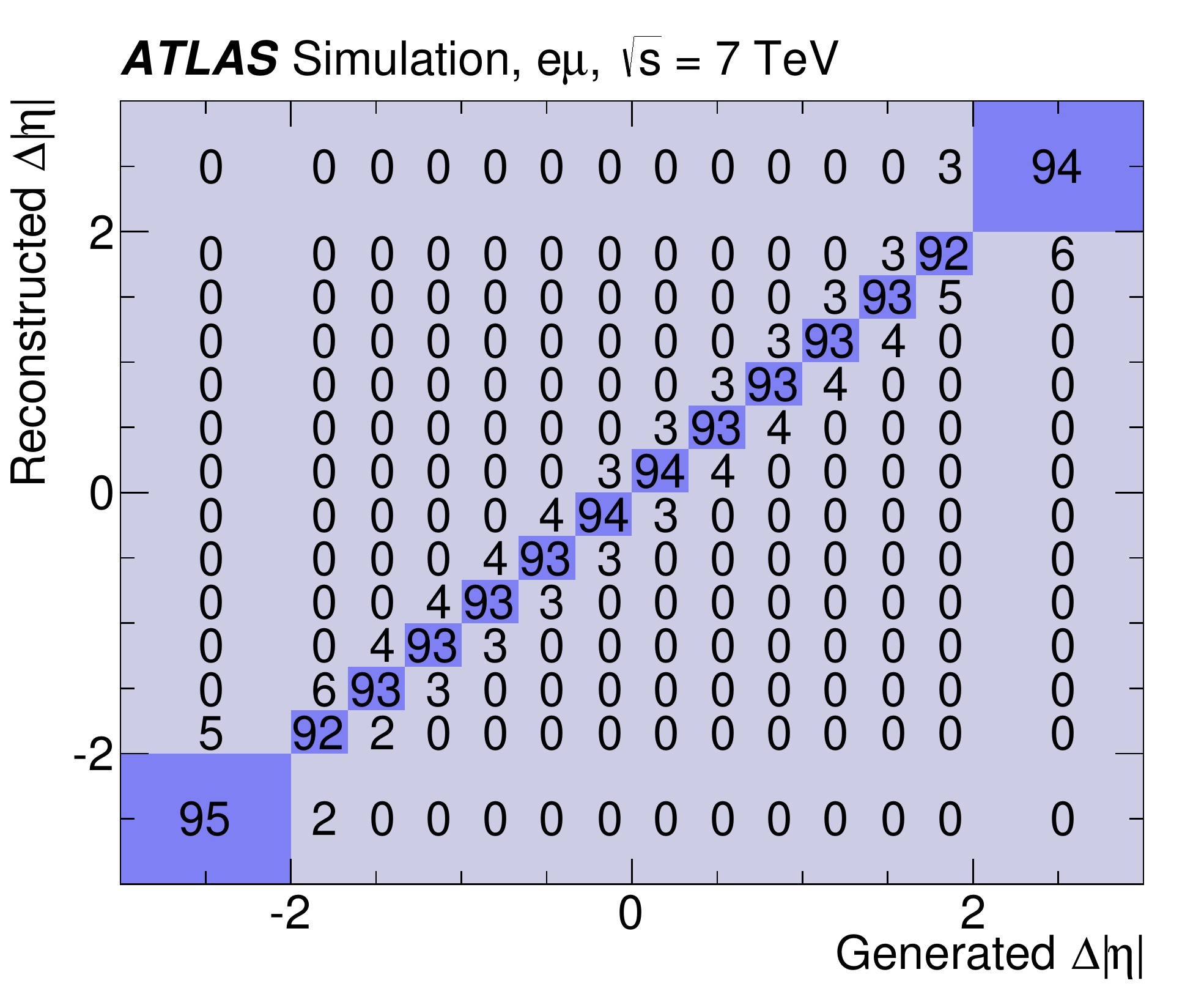}
\label{fig:mm_ll_tt:ll}
}
\subfigure[] {
\includegraphics[width=.47\textwidth]{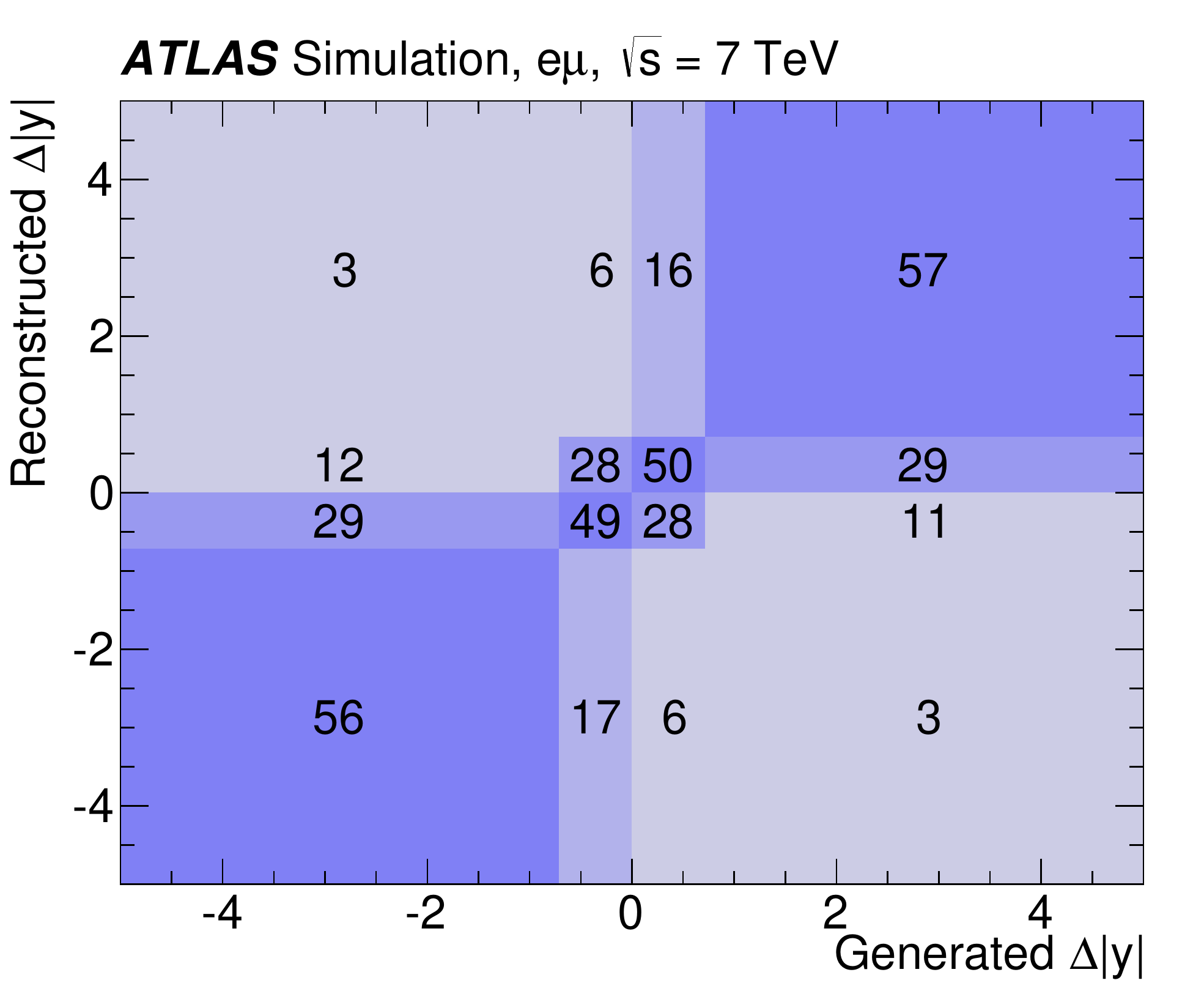}
\label{fig:mm_ll_tt:tt}
}
\caption{\label{fig:mm_ll_tt} Response matrices for (a)~the lepton \deta{} and
  (b)~\ttbar{} \dy{} observables in the \emu channel. Each column of the matrices is
    normalized to unity and values are reported as percentage (\%) units. Values smaller than 0.5\% are rounded to 0\%.}
\end{figure}

In case of the \Acll, the resolution of the measured lepton \deta{} is very
good. \FigRef{fig:mm_ll_tt:ll} shows, for the \emu{} channel, the probability of an event with a
generated value \deta{} in the $j$-th bin to be reconstructed in the $i$-th
bin of the corresponding distribution. This probability distribution is defined
to be the response matrix for the observable \deta{}. The diagonal bins of the response matrix account for more than 90\% of the events. 
The acceptance and the small migrations are accounted for by the bin-by-bin correction described in \subsecRef{subsec:bbb}.

In case of the \Actt, the top quark direction, which is necessary to determine 
the \ttbar{} asymmetry, is evaluated using the kinematic reconstruction
of the events, described in \secRef{sec:Reconstruction}. In addition to lepton
directions and energies
measured with very good resolution, jet four-momenta and \met{} measured with worse 
resolution are used in reconstructing the $t$ and $\tbar$ four-momenta. The resolution
of \ttbar{} \dy (\figRef{fig:mm_ll_tt:tt}) is thus much worse than that for the lepton
\deta. In order to correct for  detector resolution and acceptance effects in
\Actt, the fully Bayesian unfolding (FBU)
technique~\cite{Fbu2012arXiv1201.4612C} described in
\subsecRef{subsec:unfolding} is used.

\subsection{Correction of the lepton-based asymmetry}
\label{subsec:bbb}

For \Acll{}, bin-by-bin correction factors that 
also extrapolate to the full acceptance for the \ttbar{} production are used.
The goal of this procedure is to find an estimate of the true
distribution, given an observed distribution and an expected
background distribution. For the lepton-based results, true distributions are 
obtained at particle level using leptons before Quantum Electrodynamics (QED) final-state radiation\footnote{The particle-level definition uses status-code 3 for \pythia{} for electrons and muons produced in $W$ boson decays. In addition, electrons and muons produced from status-code 3 $\tau$ leptons are used. These particles are used both for the unfolding and for the predictions of the MC generators.}. 
The following notation is used: $\bm{\mu}$
and $\bm{\hat{\mu}}$ are vectors of true distribution values and its estimate,
respectively. An observed distribution is denoted by $\bm{n}$ and its expected value from
simulation by $\bm{\nu}^{\rm{MC}}$. An expected background distribution is denoted by
$\bm{\beta}$. For the $i$-th bin of the asymmetry distribution, 
the estimate of the true value is obtained by applying a correction factor $C_i$ 
to the difference between the observed number of events and the expected number of
background events,
\begin{equation}
 \hat{\mu_i}=C_i(n_i-\beta_i)~.
\label{eq:corr1}
\end{equation} 
The $C_i$ are estimated using the $\ttbar$ MC simulated sample as 
\begin{equation}
 C_i=\frac{\mu_i^{\rm{MC}}}{\nu_i^{\rm{MC}}},
\label{eq:corr2}
\end{equation}
where $\mu_i^{\rm{MC}}$ and $\nu_i^{\rm{MC}}$ are the predictions for the number
of events in the $i$-th bin of the true and reconstruction-level distributions.

The bin-by-bin correction of \Acll{} is tested on simulation samples reweighted such that different levels of asymmetry \deta\ are introduced. 
Samples are reweighted according to a linear function of \deta\ with a slope between $-6$\% and $6$\%~ in steps of $2$\%. 
Corrected values obtained from reweighted distributions are found to be in good agreement with the input value, 
following a linear relationship. 
The choice of the binning is
done by optimizing the linearity of the method and the expected statistical uncertainty of the asymmetry. 
The results in \secRef{sec:Results} are obtained with \deta\ distribution binned
in 14 bins in the interval between $-3$ and $3$.

The correction factors depend strongly on the channel and the bin, with the outer bins receiving larger fractional corrections. The \ee{} channel has the lowest acceptance and thus the highest correction factors, reaching values of 500 in the outer bins, in which the events are mostly outside the detector fiducial acceptance. The \emu{} channel has a much higher acceptance, and the correction factors vary between 10 and 60. The dependence of the correction factors on the MC model and PDF is small, up to approximately~5\%.

\subsection{Unfolding of the $\ttbar$ asymmetry}
\label{subsec:unfolding}
In case of sizeable migrations across the bins of the considered distribution,
   the migrations need to be taken into account without introducing a
   significant bias during the correction procedure. Unfolding is better suited
   for the purpose than the bin-by-bin correction factors described in
   \subsecRef{subsec:bbb}. Using the response matrix ($\bm{R}$), the true
   distribution ($\bm{\mu}$) is related to the expected reconstruction-level
   distribution ($\bm{\nu}$) and the expected background ($\bm{\beta}$) by
\begin{equation}
 \bm{\nu}=\bm{R\mu}+\bm{\beta}.
 \label{eq:unf1}
\end{equation}
In the FBU technique, the maximum likelihood estimator of $\bm{\mu}$, $L(\bm{\mu})$, is given by
\begin{equation}
 \log L(\bm{\mu})=\sum_{i=1}^{N}\log P(n_i;\nu_i)-\alpha
 S(\bm{\mu})~;\quad~p(\bm{\mu}) \propto L(\bm{\mu})~,
 \label{eq:unf3}
\end{equation}
with $P$ the Poisson distribution, $\bm{n}$ the observed distribution, $S$ a regularization function and $\alpha$ a regularization parameter. The sum in $i$ runs over all $N$ bins of the distributions.  
The probability density of the unfolded
spectra $p(\bm{\mu})$ is proportional to $L(\bm{\mu})$. The regularization
function $S$ is selected such that the spectra with a desired quality,
such as smoothness, are preferred. The regularization parameter $\alpha$ 
controls the relative strength of the regularization when
evaluating the likelihood. 
The unfolded spectrum and its associated uncertainty are extracted from the
probability density $p(\bm{\mu})$. The statistical uncertainty corresponds to
the width of the shortest interval covering 68\% probability, and the unfolded
spectrum corresponds to the middle of that interval.

The response matrix is obtained using information from the nominal $\ttbar$ simulated
sample and, in particular, using the top quarks before their decay
(parton level) and after QCD radiation\footnote{The parton-level definition uses status-code 155 for \herwig{} and 3 for \pythia for both the unfolding and for the predictions of the MC generators.}.

As explained for the lepton-based asymmetry, the correction is done at the level of true
dilepton events (where the two top quarks decay to electrons or muons, 
either from a direct $W$ boson decay or through an intermediate $\tau$ lepton decay).

Using the vector of the true distribution's estimated values $\bm{\hat{\mu}}$, the regularization function is defined based on the curvature 
$S(\bm{\mu}) = |C(\bm{\mu}) - C(\bm{\hat{\mu}})|$, with 
\begin{equation}
C(\bm{\mu}) = \sum_{i=2}^{N -1} (\Delta_{i+1,i} - \Delta_{i,i-1})^2, \quad
\textrm{and } \quad \Delta_{i+1,i} = \mu_{i+1} - \mu_{i}.
\end{equation}
As in the case of the lepton-based asymmetry, linearity tests are 
performed. A given asymmetry value is introduced by reweighting the samples according to a linear function of \ttbar{} \dy\ with a slope between -6\% and 6\%~ in steps of 2\%. Unfolded values obtained from reweighted distributions are observed to be in good agreement with the injected values of \Actt, following a linear relationship. 
This linearity test is performed with and without regularization and yields
similar performance. 
The binning used for the \dy distribution as well as the regularization
parameter are optimized simultaneously to minimise the expected 
statistical uncertainty while achieving good linearity. 
The results in \secRef{sec:Results} are obtained with a regularization parameter
$\alpha=10^{-7}$. The \dy\ distribution is binned in 4 bins in the interval
between $-5$ and $5$. 
For this binning choice, at least 50\% of the events populate the response matrix diagonal bins for each of the \ee, \emu and \mumu channels (\figRef{fig:mm_ll_tt:tt}).

The overall correction which is applied to the distribution varies between
factors of 10 and 100, depending on the channel and the bin. As shown in
\figRef{fig:mm_ll_tt} the bins used for the \ttbar{} \dy{} distribution are
wider than the bins used for the lepton \deta\ distribution. The acceptance
correction applied to the outer bins of \dy{} is thus smaller than the
correction obtained for the outer bins of \deta\ distribution.

\section{Systematic uncertainties}
\label{sec:Systematics}
The uncertainties of the \Acll{} corrections and \Actt{} unfolding method are estimated 
from the non-closure in the linearity test described in~\secRef{sec:Corrections}. 
For \Acll{} the uncertainties are $-0.002$ in \ee{} channel and negligible ($<0.001$) in 
the \emu{} and \mumu{} channels. 
For \Actt{} the uncertainties are $0.002$ in the \ee{} channel and negligible ($<0.001$) 
in the \emu{} and \mumu{} channels. 
For both \Acll{} and \Actt{} the uncertainties have a negligible contribution to the measurement 
uncertainty and are not considered for the evaluation of the total systematic uncertainty of the results.

The systematic uncertainties considered in this analysis are classified into
three categories: detector modelling uncertainties, signal modelling
uncertainties and uncertainties related to the estimation of the backgrounds. The contributions of these sources of uncertainty are summarized in \tabRef{tab:ll_sys_comb} for the lepton-based asymmetry \Acll{} and in \tabRef{tab:tt_sys_comb} for the \ttbar{} asymmetry \Actt{}. The resulting variations are assumed to be of the same size in both directions and are therefore symmetrized. Apart from one-sided uncertainties, as in the case of the comparison of different MC models, the symmetrization does not notably modify the uncertainty values. 

\begin{table}[tbp]
\centering
\begin{tabular}[!ht]{|l|r|r|r|r|}
\hline
                      & \multicolumn{1}{|c|}{ $ee$}  & \multicolumn{1}{|c|}{ $e\mu$} & \multicolumn{1}{|c|}{$\mu\mu$} & \multicolumn{1}{|c|}{ comb.}\\ 
\hline
{\bf Measured value}         & 0.101      & 0.009      & 0.047      & 0.024 \\
{\bf Statistical uncertainty}& $\pm$0.052 & $\pm$0.019 & $\pm$0.030 & $\pm$0.015 \\
\hline
Lepton reconstruction       & $\pm$0.011 & $\pm$0.008 & $\pm$0.009   & $\pm$0.008 \\[1ex]
Jet reconstruction          & $\pm$0.006 & $\pm$0.001 & $\pm$0.004   & $\pm$0.001 \\[1ex]
\met{}                      & $\pm$0.001 & $<$0.001   & $\pm$0.002   & $<$0.001   \\[1ex]
Signal modelling            & $\pm$0.004 & $\pm$0.003 & $\pm$0.003   & $\pm$0.003 \\[1ex]
PDF                         & $\pm$0.004 & $<$0.001   & $<$0.001     & $<$0.001   \\[1ex]
NP \& fake                  & $\pm$0.016 & $\pm$0.001 & $\pm$0.001   & $\pm$0.001 \\[1ex]
Background                  & $\pm$0.003 & $\pm$0.002 & $<$0.001     & $\pm$0.001 \\[1ex]
\hline
{\bf Total sys.}            & $\pm$0.021 & $\pm$0.009 & $\pm$0.012 & $\pm$0.009 \\
\hline
\end{tabular}

\caption{\label{tab:ll_sys_comb}  Measured value and uncertainties for the
  lepton-based asymmetry \Acll. Uncertainties with absolute value below 0.001
    are considered negligible for the total uncertainty.}
\end{table}

\begin{table}[tbp]
\centering
\begin{tabular}[!ht]{|l|r|r|r|r|}
\hline
                      & \multicolumn{1}{|c|}{ $ee$} & \multicolumn{1}{|c|}{ $e\mu$} & \multicolumn{1}{|c|}{ $\mu\mu$} & \multicolumn{1}{|c|}{ comb.}   \\
\hline
{\bf Measured value}          & 0.025      & 0.007      & 0.043      & 0.021      \\
{\bf Statistical uncertainty} & $\pm$0.069 & $\pm$0.032 & $\pm$0.045 & $\pm$0.025 \\
\hline
Lepton reconstruction       & $\pm$0.008 & $\pm$0.008 & $\pm$0.004     & $\pm$0.007    \\[1ex]
Jet reconstruction          & $\pm$0.015 & $\pm$0.009 & $\pm$0.006     & $\pm$0.009    \\[1ex]
\met{}                      & $\pm$0.015 & $\pm$0.005 & $\pm$0.008     & $\pm$0.007    \\[1ex]
Signal modelling            & $\pm$0.004 & $\pm$0.004 & $\pm$0.003     & $\pm$0.003    \\[1ex]
PDF                         & $\pm$0.004 & $\pm$0.005 & $\pm$0.004     & $\pm$0.005    \\[1ex]
NP \& fake                  & $\pm$0.013 & $\pm$0.011 & $\pm$0.003     & $\pm$0.008    \\[1ex]
Background                  & $\pm$0.003 & $\pm$0.002 & $\pm$0.004     & $\pm$0.003    \\[1ex]
\hline
{\bf Total sys.}            & $\pm$0.027 & $\pm$0.018 & $\pm$0.013 & $\pm$0.017 \\
\hline
\end{tabular}

\caption{\label{tab:tt_sys_comb} Measured value and uncertainties for the \ttbar{} asymmetry \Actt.}
\end{table}

Detector modelling uncertainties are evaluated by performing corrections for
detector effects for \Acll and \Actt, with the response matrices
corresponding to the systematic variations. Effects of detector modelling
uncertainties on the background are included by subtracting the 
background, varied accordingly, from the data. The following sources are considered.
\begin{itemize}
\item {\bf Lepton reconstruction}\\
The uncertainty due to lepton reconstruction includes several sources. Lepton momentum scale and resolution modelling correction factors and associated uncertainties are derived from
comparisons of data and simulation in $Z\ra\ll$ events~\cite{Aad:2014fxa, Aad:2014zya}. 
Uncertainties in the modelling of trigger, reconstruction and lepton identification efficiencies are
also included. Data-to-simulation efficiency corrections, and their uncertainties, are
derived from $J/\psi\ra \ll$, $Z\ra\ll$ and $W\ra e\nu$ events. 
\item {\bf Jet reconstruction}\\
The effects include the jet energy scale and jet
resolution uncertainties. Jet energy scale uncertainty is derived using
information from test-beam data, LHC collision data and
simulation~\cite{Aad:2014bia}. It includes uncertainties in the flavour
composition of the samples and mismeasurements due to close-by jets and pileup effects. Jet energy resolution and reconstruction efficiency uncertainties are obtained using minimum bias and QCD dijet events~\cite{Aad:2014bia,Aad:2012ag}. 
\item {\boldmath \met{}}\\ 
The uncertainties from the energy scale and resolution
corrections for leptons and jets are propagated to the \met. The category
accounts for uncertainties in the energies of calorimeter cells not associated with the reconstructed objects and the uncertainties from cells associated with low-\pt{} jets (7\,\GeV $<$ \pT $<$ 20\,\GeV)~\cite{Aad:2012re} as well as the dependence of their energy on the number of pileup interactions. 
\end{itemize}

The uncertainties due to the modelling of the signal \ttbar{} distributions are evaluated by 
performing the linearity test for signal model samples generated with various assumptions. The following sources are quoted.
\begin{itemize}
\item {\bf Signal modelling}\\
The uncertainty is evaluated by adding in
quadrature the
MC generator uncertainties, initial- and final-state radiation (ISR and FSR), underlying event (UE) and colour reconnection (CR) uncertainties described in the following. 
The systematic uncertainty related to the choice of a MC generator 
includes the difference between the nominal sample generated with \powheg{} + \pythia and 
samples generated with \mcatnlo{} + \herwig, \powheg{} + \herwig and \alpgen{} + \herwig.
The effects of renormalization and factorization scale choice are evaluated with
a dedicated pair of samples generated with \mcatnlo{} + \herwig. In these samples
renormalization and factorization scales are varied simultaneously by a factor
of two with respect to the reference scale. The reference scale is fixed at the
\mcatnlo{} generator default, which is defined as the average of the $t$ and the
$\bar{t}$ transverse masses, $Q=\sqrt{1/2({\pt}_{t}^2 + {\pt}_{\tbar}^2) +
  \mt^2}$, where ${\pt}_{t(\tbar)}$ corresponds to the transverse momentum of
  the $t$ or $\tbar$. Since the effects covered by generator comparisons and
  scale variations partially overlap, only the largest contribution from all
  comparisons is used. For the lepton-based asymmetry the dominant contribution was found to stem from 
the difference between the nominal sample and the sample generated with \alpgen{} + \herwig. For the
\ttbar{} asymmetry the contributions of the comparison of the baseline sample result to the results obtained with each of  \mcatnlo{} + \herwig, \powheg{} + \herwig{} and \alpgen{} + \herwig{} samples are of comparable size and significantly larger than the contribution of the renormalization and factorization scale uncertainty.
The amount of ISR and FSR are treated as an additional source of signal
  modelling uncertainty. It is evaluated using samples generated with
  \alpgen{} + \pythia{} with variations of parameters controlling the
  renormalization scale used in \alpgen{} and in the \pythia parton shower. The
  renormalization scale is varied by factors of 0.5 and 2.
The \pythia{} settings correspond to Perugia radLO and radHi
tunes~\cite{PhysRevD.82.074018}. Apart from this, the UE and CR uncertainties are evaluated by comparing samples generated with 
\powheg{} + \pythia, using Perugia2011, Perugia2011 mpiHi and Perugia2011 noCR
tunes~\cite{PhysRevD.82.074018}. For \Acll{}, the contributions from the choice
of MC generator and from ISR and FSR exceed the non-perturbative UE and CR
contributions. For \Actt{}, the contributions from the choice of MC generator 
and from radiation and non-perturbative modelling uncertainties are comparable. 
\item {\bf PDF uncertainty}\\
The uncertainty due to the PDF is evaluated by performing linearity tests with samples obtained from the nominal signal sample, generated with CT10 PDF, reweighted to other PDFs. 
The CT10 error set as well as MSTW2008 68\% CL NLO~\cite{Martin:2009iq} and NNPDF2.3 NLO ($\alpha_{\rm{s}}=0.118$)~\cite{Ball:2012cx} central predictions are used. 
For each asymmetry value, the largest value of the three sources is quoted as
uncertainty.
\end{itemize}

The uncertainties on the modelling of the SM backgrounds are divided into two
categories described below.
\begin{itemize}
\item {\bf NP \& fake}\\ 
This source corresponds to the uncertainty 
in the estimation of processes fulfilling the event selection due to non-prompt or misidentified 
leptons. 
The uncertainties are obtained by varying the efficiencies for a real or fake
lepton to pass the tight selection, and are affecting both the normalization of the
background and its shape. 
\item {\bf Background}\\
The uncertainties in the modelling of diboson, $Z$+jets and single-top SM
processes are quoted in the background category. They are evaluated by varying the
normalization of each of these processes by the uncertainty on its
cross-section. The uncertainty on the overall luminosity of 1.8\% is also
entering this category~\cite{Aad:2013ucp}.
\end{itemize} 

For both the lepton-based asymmetry \Acll and the \ttbar{} asymmetry \Actt, the
statistical uncertainty is larger than the total systematic uncertainty. The
\Acll{} measurement has a combined statistical uncertainty of 1.5\%, whereas the
combined systematic uncertainty is 0.9\%. The largest source of \Acll{}
systematic uncertainty is the lepton reconstruction uncertainty, which accounts
for approximately 90\% of the total systematic uncertainty. The uncertainty on
the asymmetry \Acll{} measured
in the \ee{} channel receives a sizeable contribution from the NP \& fake leptons
category (1.6\%). This, however, does not significantly impact the combined systematic
uncertainty since the \ee{} channel receives a small weight in the
combination, as detailed in~\secRef{sec:Results}. The \ttbar{} asymmetry
\Actt{} has a combined statistical uncertainty of 2.5\% and a combined systematic
uncertainty of 1.7\%. The detector modelling uncertainties account for
approximately 80\% of the combined systematic uncertainty, 
with comparable large contributions from the lepton reconstruction, the \met{}
(0.7\%) and the jet reconstruction uncertainty (0.9\%). 
The NP \& fake contribution to the \Actt{} systematic
uncertainty is also sizeable (0.8\%).

The uncertainties related to detector and background modelling are evaluated in each bin of the corrected distributions and presented in \secRef{sec:Results}.

\section{Results}
\label{sec:Results}
After the event selection and reconstruction but before the correction described in~\secRef{sec:Corrections} the inclusive lepton and \ttbar{} asymmetries measured in the data are \Acll = 0.021 $\pm$ 0.011 (stat.) and \Actt = 0.003 $\pm$ 0.012 (stat.), respectively for the combination of the \ee,\emu{} and \mumu{} channels. After the subtraction of the background contribution, the measured data asymmetries are \Acll = 0.029 $\pm$ 0.013 (stat.) and \Actt = 0.006 $\pm$ 0.014 (stat.). The corresponding asymmetry predictions in the nominal simulated \ttbar{} sample are \Acll = 0.005 $\pm$ 0.003 (stat.) and \Actt = 0.008 $\pm$ 0.003 (stat.). 
This sample is generated with the \powheg{} + \pythia{} generator with a particle-level lepton asymmetry of \Acll = \Acllpowheg{} and a parton-level \ttbar{} asymmetry of \Actt = \Acttpowheg{}, evaluated in the full phase-space.

After the correction for detector, resolution and acceptance effects, the
normalized differential cross-sections corrected to particle and 
parton level are obtained for \deta{} and \dy{} 
separately for the three channels.
From these distributions, the inclusive asymmetry values can be extracted.
The inclusive results obtained in the \ee, \emu{} and \mumu{} channels (see \tabsAndTabRef{tab:ll_sys_comb}{tab:tt_sys_comb}) 
are then combined 
using the best linear unbiased estimator (BLUE) method~\cite{Lyons:1988rp, Valassi:2003mu}.
All systematic uncertainties are assumed to be 100\% correlated, except for the
uncertainties on electrons and muons and on the NP \& fake lepton background. 

\begin{figure}[tbp]
\centering 
\subfigure[] {
\includegraphics[width=.47\textwidth]{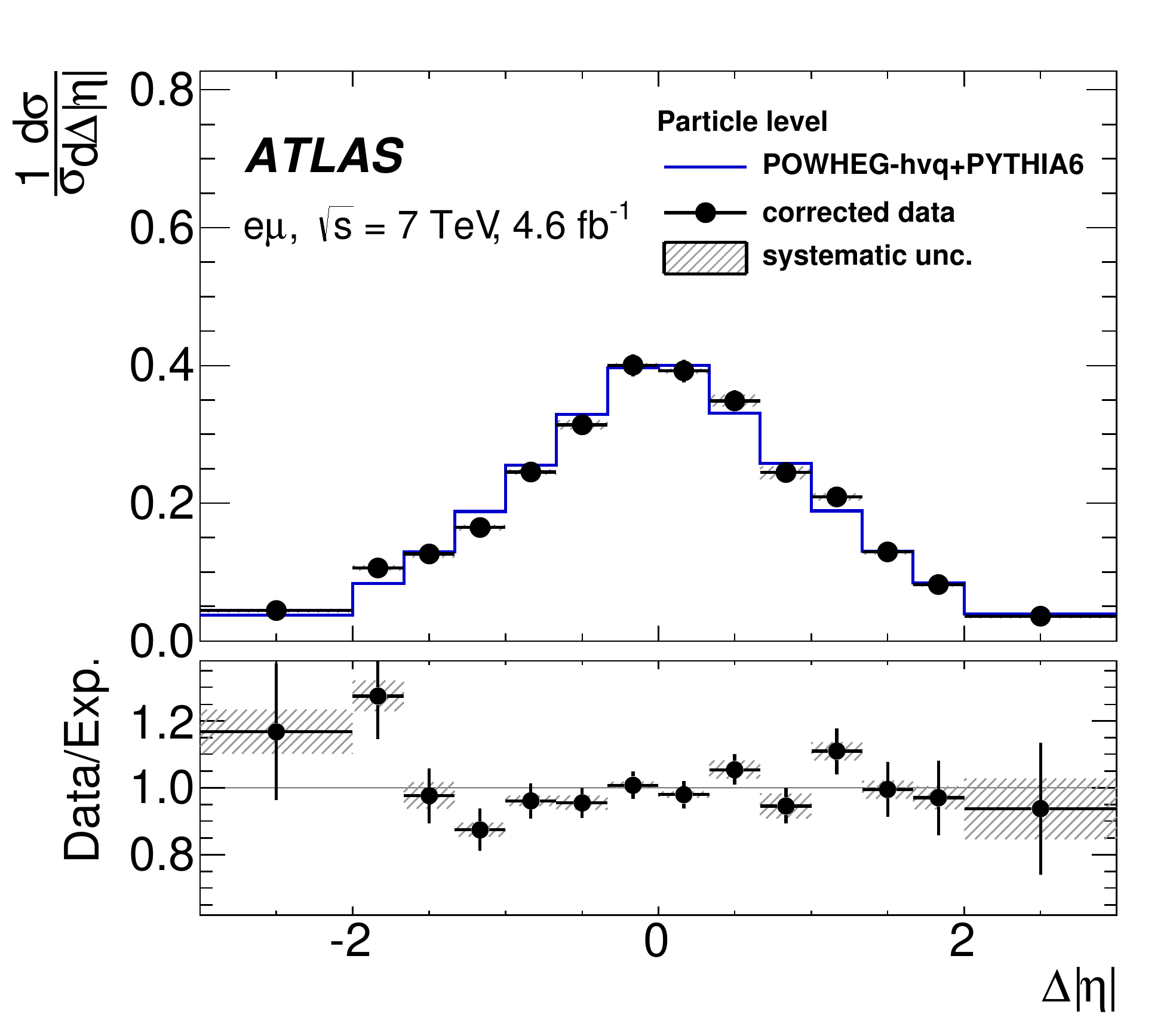}
\label{fig:ll_corr_spectra}
}
\subfigure[] {
\includegraphics[width=.47\textwidth]{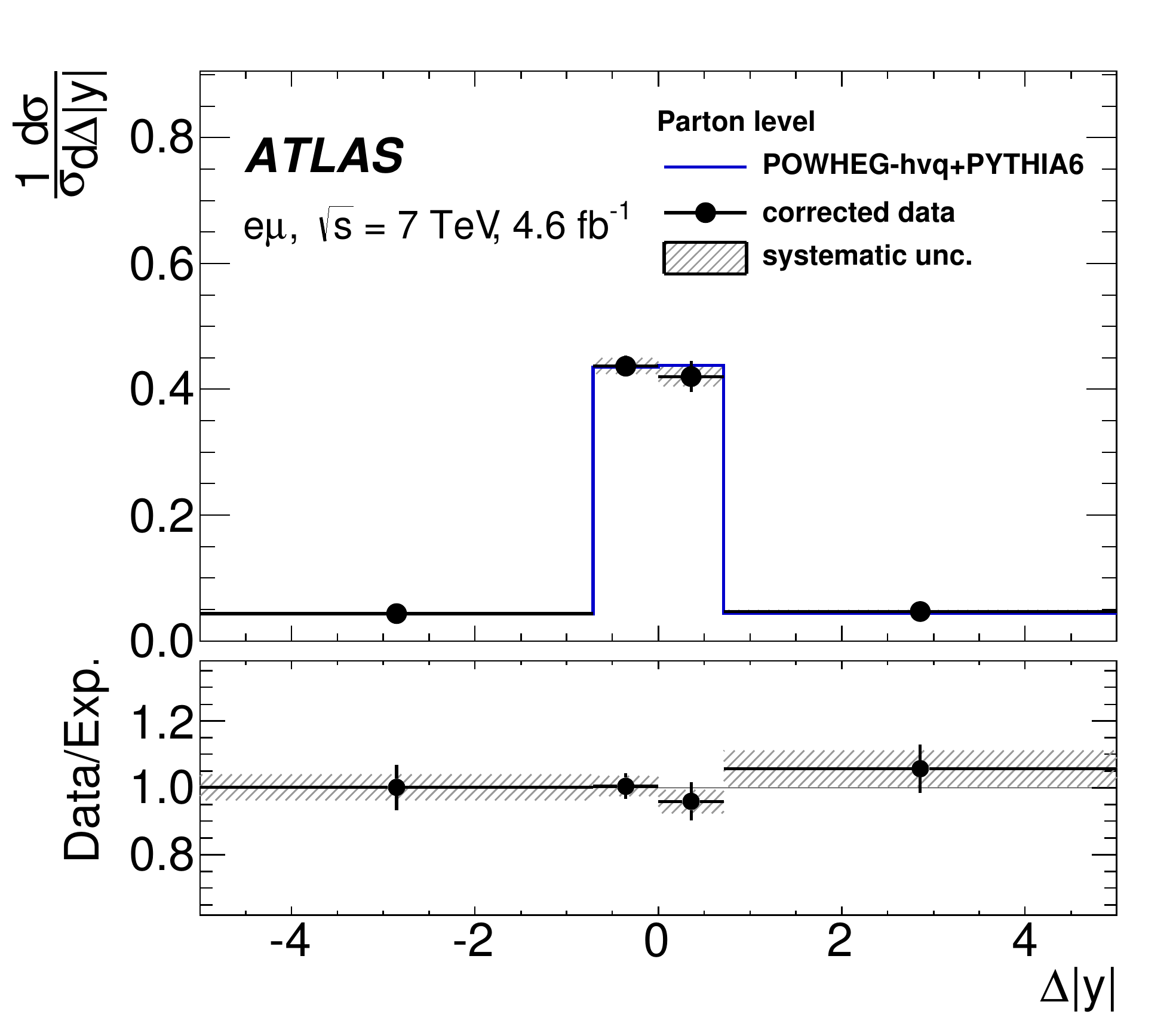}
\label{fig:tt_corr_spectra}
}
\caption{\label{fig:corr_spectra} Normalized differential
cross-sections for (a) lepton \deta{} and (b) \ttbar{} \dy{} 
in the \emu{} channel after correcting for detector effects. 
The distributions predicted by \powheg{} + \pythia{} are compared to the data in the top
panels. The bottom panels show the ratio of the corrected data to
the predictions.
The error bars correspond to the statistical uncertainties and the hatched area to the systematic uncertainties.}
\end{figure}

The normalized differential cross-sections for \deta{} and \dy{} 
are presented in~\figRef{fig:corr_spectra} for the \emu{} channel. 
Good agreement is observed between the measured distributions and the ones predicted by \powheg{} + \pythia.
The normalized differential cross-sections in that channel are also presented with 
statistical and systematic uncertainties in
\tabsAndTabRef{tab:diffxsec_leptons}{tab:diffxsec_ttbar}. 
The systematic uncertainties for the differential distributions do not include the 
signal modelling uncertainties, which could not be evaluated with sufficient precision 
due to the limited statistics of the simulated samples. 
For both distributions, the statistical uncertainty is somewhat larger 
than the systematic uncertainty. 
In \appRef{app:Appendix} the contributions 
from each source of systematic uncertainty, described in \secRef{sec:Systematics}, 
to the total systematic uncertainty in each bin of the \deta{} and \dy{} distributions 
are provided. The statistical correlations between the different bins of the distributions are also given. 
 
\begin{table}[htbp]
  \begin{center}
  \begin{tabular}{| c | rll|}
  \hline
  Bin of \deta & \multicolumn{3}{c|}{$ \frac{1}{\sigma} \frac{d\sigma}{d\Delta|\eta|}$ ($\pm$ stat. $\pm$ syst.)} \\
    \hline
      $[-3.00,-2.00]$   & 0.0440 & $\pm$ 0.0077 & $\pm$ 0.0025 \\
      $[-2.00,-1.67]$   & 0.106  & $\pm$ 0.011  & $\pm$ 0.004  \\
      $[-1.67,-1.33]$   & 0.126  & $\pm$ 0.011  & $\pm$ 0.005  \\
      $[-1.33,-1.00]$   & 0.164  & $\pm$ 0.012  & $\pm$ 0.004  \\
      $[-1.00,-0.67]$   & 0.245  & $\pm$ 0.013  & $\pm$ 0.004  \\
      $[-0.67,-0.33]$   & 0.314  & $\pm$ 0.015  & $\pm$ 0.007  \\
      $[-0.33,0.00]$    & 0.400  & $\pm$ 0.016  & $\pm$ 0.004  \\
      $[0.00,0.33]$     & 0.392  & $\pm$ 0.016  & $\pm$ 0.004  \\
      $[0.33,0.67]$     & 0.349  & $\pm$ 0.015  & $\pm$ 0.009  \\
      $[0.67,1.00]$     & 0.244  & $\pm$ 0.013  & $\pm$ 0.010  \\
      $[1.00,1.33]$     & 0.209  & $\pm$ 0.013  & $\pm$ 0.005  \\
      $[1.33,1.67]$     & 0.129  & $\pm$ 0.011  & $\pm$ 0.003  \\
      $[1.67,2.00]$     & 0.0815 & $\pm$ 0.0093 & $\pm$ 0.0028  \\
      $[2.00,3.00]$     & 0.0361 & $\pm$ 0.0076 & $\pm$ 0.0035  \\
    \hline
    \end{tabular}
  \caption{Normalized differential cross-sections for \deta in the \emu{} channel presented with
    statistical and systematic uncertainties.}
  \label{tab:diffxsec_leptons}
  \end{center}
\end{table}

\begin{table}[htbp]
  \begin{center}
  \begin{tabular}{| c | rll |}
  \hline
  Bin of \dy & \multicolumn{3}{c|}{$ \frac{1}{\sigma} \frac{d\sigma}{d\Delta|y|}$ ($\pm$ stat. $\pm$ syst.)} \\
    \hline
      $[-5.00, -0.71]$ & 0.0435 & $\pm$ 0.0029 & $\pm$ 0.0017 \\
      $[-0.71, 0.00]$  & 0.437  & $\pm$ 0.016  & $\pm$ 0.013  \\
      $[0.00, 0.71]$   & 0.420  & $\pm$ 0.025  & $\pm$ 0.015  \\
      $[0.71, 5.00]$   & 0.0470 & $\pm$ 0.0032 & $\pm$ 0.0024 \\
    \hline
    \end{tabular}
  \caption{Normalized differential cross-sections for \dy in the \emu{} channel presented with
    statistical and systematic uncertainties.}
  \label{tab:diffxsec_ttbar}
  \end{center}
\end{table}

The results for the inclusive lepton-based asymmetry \Acll{} and the \ttbar{} asymmetry \Actt{} 
after corrections for detector and resolution effects are shown in
\tabRef{tab:ac_unfolded}. The values in the \ee, \emu{} and \mumu{} channels as well
as for their combination are presented, together with statistical and
systematic uncertainties. 

\begin{table}[tbp]
\centering

\begin{tabular}{|l|c|c|}
\hline
Channel &  \Acll & \Actt \\
\hline
\ee   & 0.101  $\pm$ 0.052 $\pm$ 0.021 & 0.025 $\pm$ 0.069 $\pm$  0.027 \\
\emu  & 0.009  $\pm$ 0.019 $\pm$ 0.009 & 0.007 $\pm$ 0.032 $\pm$  0.018 \\
\mumu & 0.047  $\pm$ 0.030 $\pm$ 0.012 & 0.043 $\pm$ 0.045 $\pm$  0.013 \\
\hline
Combined & \Acllcombtab & \Acttcombtab \\
\hline
SM, NLO QCD+EW~\cite{Bernreuther:2012sx} & \Aclltheory & \Actttheory \\
\hline         
\end{tabular}

\caption{\label{tab:ac_unfolded} Results for the lepton-based asymmetry 
  \Acll and the \ttbar asymmetry \Actt after correcting for detector, resolution and acceptance effects. 
  The values in the \ee, \emu and \mumu channels as well as the combined value are presented with their 
  statistical and systematic uncertainties.}
\end{table}

Detailed information about the combination of the inclusive values is given in \tabRef{tab:combdetails}.
The combination probabilities are 21\% and 81\% for \Acll and \Actt respectively, 
demonstrating the compatibility of the measurements in the three channels (\ee{}, \emu{} and \mumu{}).
The weight of each channel in the combination is also reported in \tabRef{tab:combdetails}. 
The \emu{} channel dominates the combination, reflecting the larger data statistics compared to that of the \ee{} and \mumu{} channels.

\begin{table}[tbp]
 \centering
   \begin{tabular}[!ht]{|l|c|c|}
   \hline
                                            & \Acll              & \Actt \\
    \hline
    $\chi^2$                                & 3.1                & 0.4 \\
    Probability (in \%)                     & 21                 & 81 \\
    Weights ($ee$/$e\mu$/$\mu\mu$ in \%)    & 7 / 68 / 25        & 9 / 57 / 34   \\
    \hline
  \end{tabular}

  \caption{Information about the combination of the three channels using the best linear unbiased estimator method:
    $\chi^2$ and probability of the combination, as well as the weight of each
      channel.}
 \label{tab:combdetails}
\end{table}

\begin{figure}[tbp]
\centering 
\includegraphics[width=.47\textwidth]{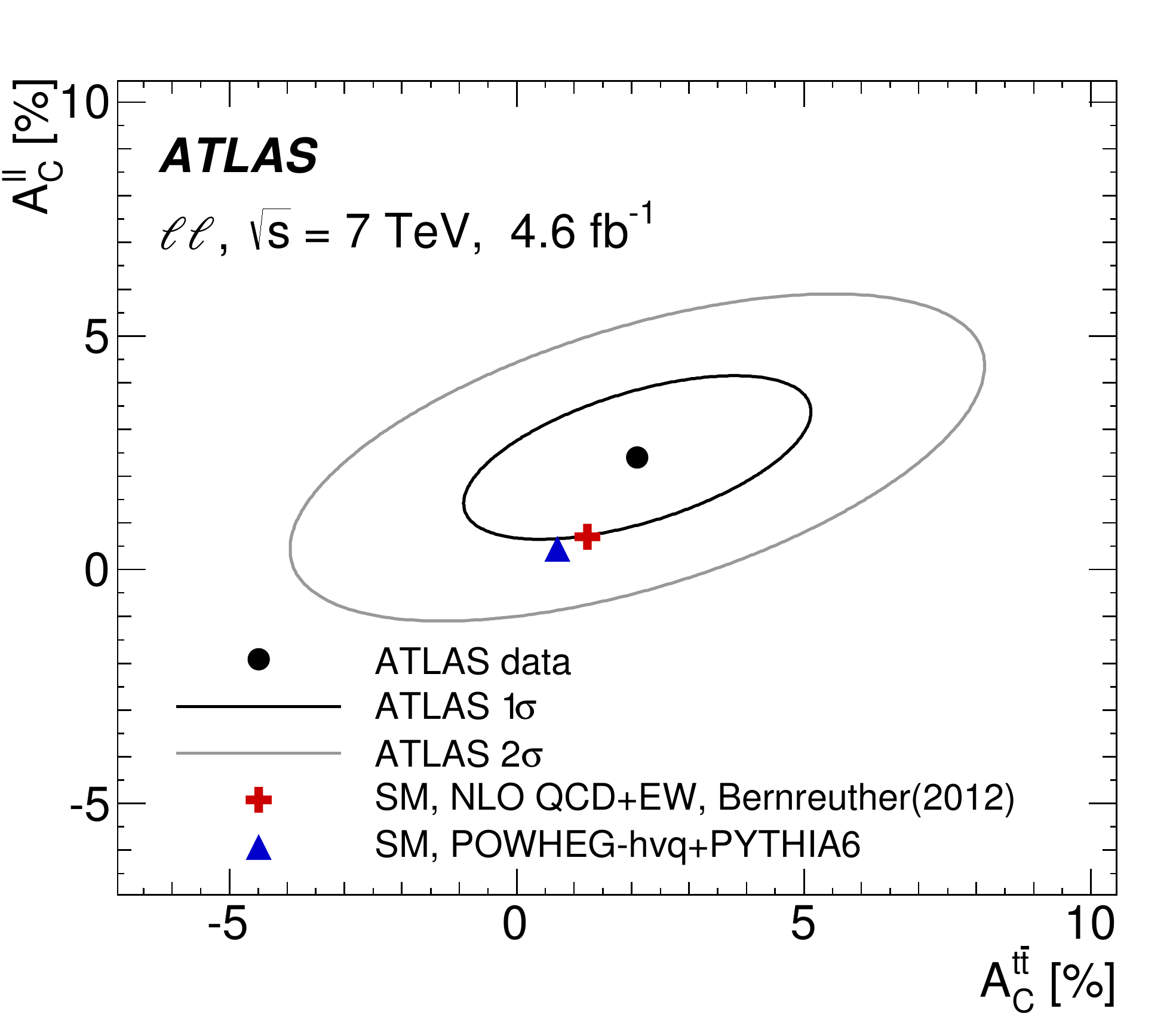}
\caption{\label{fig:acll_actt_twod} Comparison of the inclusive \Acll and \Actt measurement values to the theory predictions 
  (SM NLO QCD+EW prediction~\cite{Bernreuther:2012sx} and the prediction of the \powheg{} + \pythia generator). 
  Ellipses corresponding to 1$\sigma$ and 2$\sigma$ combined statistical and systematic uncertainties of the measurement, including the correlation between \Acll and \Actt, are also shown.}
\end{figure}

The inclusive measurements after the detector and resolution effects corrections can be compared with the state-of-the-art theoretical predictions calculated at NLO QCD, including the electromagnetic and weak-interaction corrections~\cite{Bernreuther:2012sx}: $\Acll = \Aclltheory$ and $\Actt = \Actttheory$. In~\figRef{fig:acll_actt_twod} the measured values of \Acll and \Actt are compared to these predictions and \powheg{} + \pythia{} predictions. In the figure, ellipses corresponding to 1$\sigma$ and 2$\sigma$ combined statistical and systematic uncertainties of the measurement, including the correlation between \Acll and \Actt, are also shown. 
The statistical correlation between \Acll and \Actt is evaluated to be 37$\pm$5\% using pseudo-experiments based on simulation.
The systematic uncertainties are treated as 100\% correlated. 
The resulting correlation between \Acll and \Actt is about 55\%. The measured values are both consistent with the theory predictions within the uncertainties. The measured \Actt values are consistent with but less precise than measurements in the single-lepton decay channel by the ATLAS~\cite{Aad:2013cea} and CMS~\cite{Chatrchyan:2012xv} collaborations. The measurements of \Acll and \Actt are also consistent with the CMS collaboration measurements in the dilepton decay channel~\cite{Chatrchyan:2014yta}.

\begin{figure}[tbp]
\centering 
\subfigure[] {
\includegraphics[width=.47\textwidth]{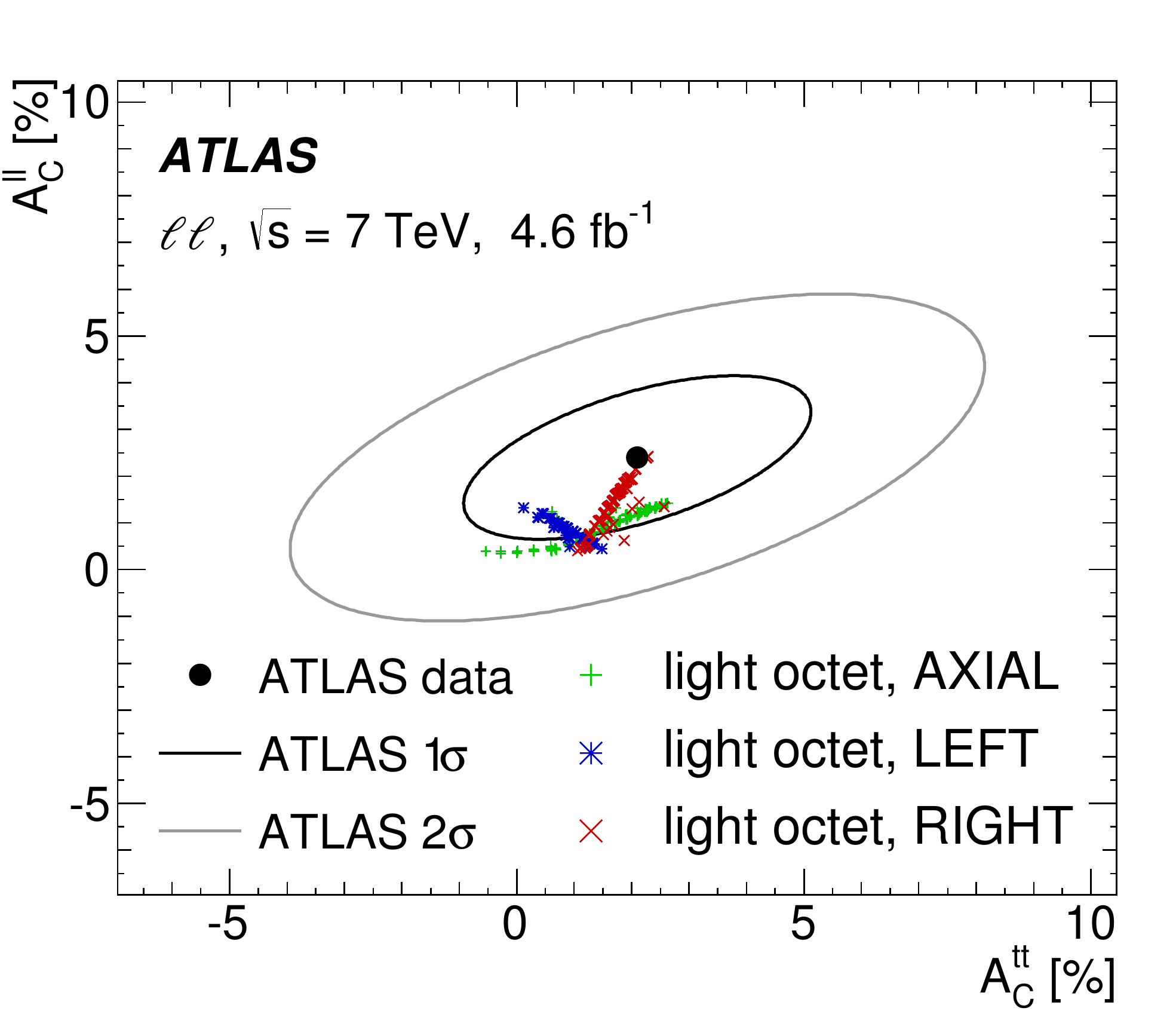}
\label{fig:acll_actt_twod_bsm_light}
}
\subfigure[] {
\includegraphics[width=.47\textwidth]{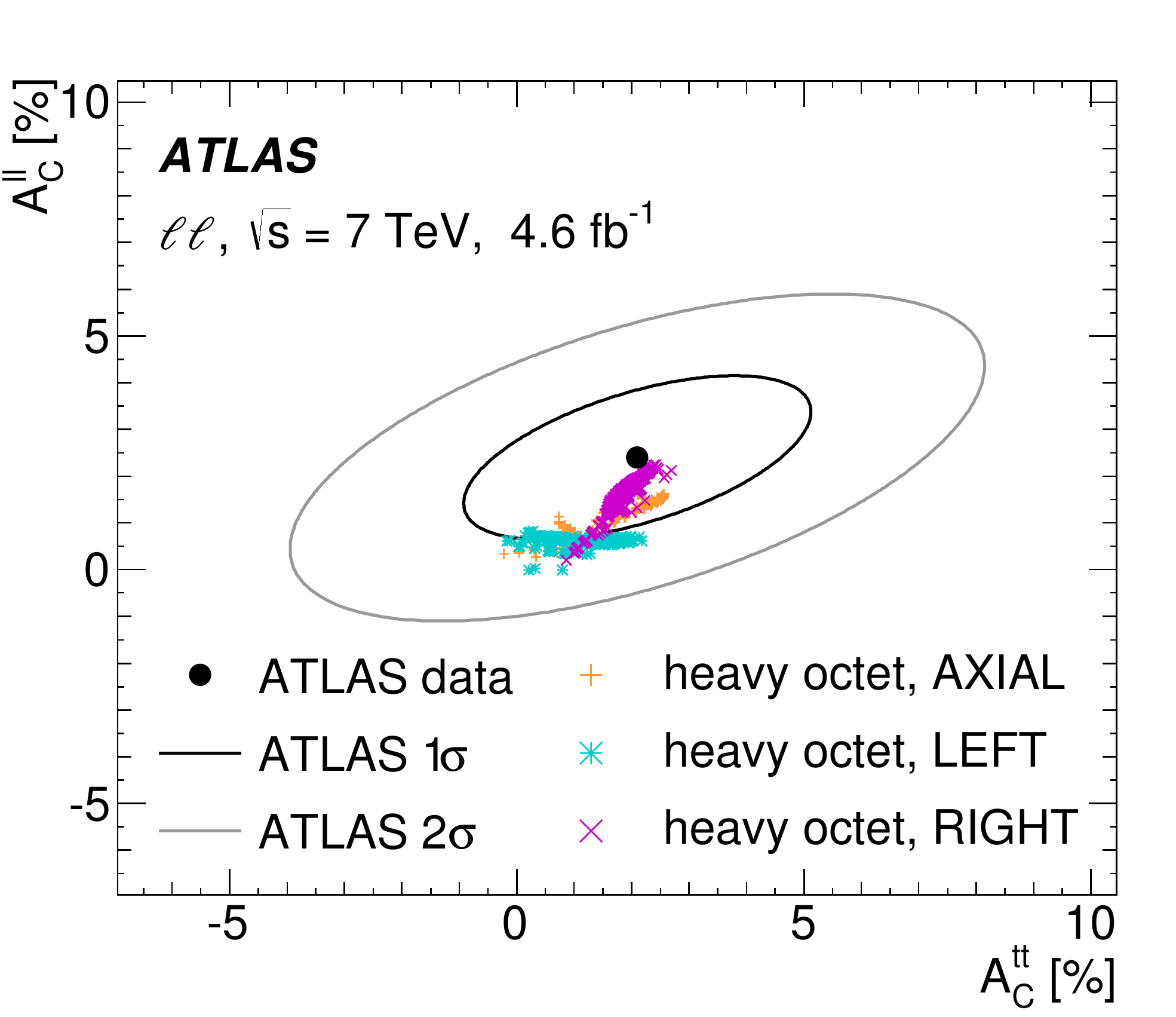}
\label{fig:acll_actt_twod_bsm_heavy}
}
\caption{\label{fig:acll_actt_twod_bsm} 
 Comparison of the measured inclusive \Acll and \Actt values to two benchmark BSM models, one a light octet with mass below \ttbar{} production threshold (left) and one with a heavy octet with mass beyond LHC reach (right), for various couplings as described in the legend.}
\end{figure}

The inclusive measurement of \Acll{} and \Actt{} is furthermore compared to two models of physics beyond the Standard Model (BSM)~\cite{Aguilar-Saavedra:2014nja} that could be invoked to explain an anomalous forward-backward asymmetry at the Tevatron, such as reported by the CDF experiment~\cite{CDF2}. Two models with a new colour octet particle exchanged in the s-channel are considered. In the model with the light octet, the new particle mass is below the \ttbar{} production threshold. The model with the heavy octet uses the octet mass beyond the reach of the LHC. The new particles would not be visible as resonances in the \mtt{} spectrum at the Tevatron or at the LHC. The light octet is assumed to have a mass of $m=250$\,\GeV{} and a width of $\Gamma=0.2m$. For the heavy octet, the corrections to \ttbar{} production are independent of the mass but instead depend on the ratio of coupling to mass, which is assumed to be 1/\TeV. In \figRef{fig:acll_actt_twod_bsm} the measured \Acll{} and \Actt{} values are compared to the light (\figRef{fig:acll_actt_twod_bsm_light}) and heavy (\figRef{fig:acll_actt_twod_bsm_heavy}) colour octet model predictions in order to assess whether any of the BSM predictions can be excluded. Models with left-handed, right-handed and axial coupling to the up, down and top quarks are shown. The considered couplings to the quarks are such that the global fit to \ttbar{} observables at the Tevatron and the LHC, including total cross-sections, various asymmetries, the top polarisation and spin correlations, is consistent with the measurements within two standard deviations~\cite{Aguilar-Saavedra:2014nja}. The LHC asymmetry measurements in the dilepton decay channel are excluded from the fit. While the models span a sizeable range of values in the \Acll{} and \Actt{} plane in~\figRef{fig:acll_actt_twod_bsm}, their predictions are consistent with the measured value within the present uncertainties. Thus the potential BSM contributions cannot be excluded beyond the reach of the previous Tevatron and LHC measurements. Future \Acll{} and \Actt{} measurements with a larger dataset could however further constrain the allowed couplings of the colour octet models if both statistical and systematic uncertainties can be reduced further.

\section{Conclusion}
\label{sec:Conclusion}
Measurements of the \ttbar{} charge asymmetry in the dilepton channel are
presented. The measurements are performed   
using data corresponding to an integrated luminosity of \lumi of \pp collisions 
at $\sqrt{s} = 7$~\tev\  collected by the \atlas detector at the LHC. 
Selected events are required to have exactly two charged leptons (electron or muon), 
large missing transverse momentum and at least two jets. Both the lepton-based asymmetry 
\Acll and the \ttbar asymmetry \Actt are extracted in three channels: \ee, \emu
and \mumu. 
The measurement of \Actt
requires the kinematic reconstruction of the \ttbar system, which is performed using
the neutrino weighting technique. Agreement between predictions and data is checked after selection and
kinematic reconstruction. Good agreement is obtained for all the kinematic
observables studied.
The $\ll$ \deta and \ttbar \dy distributions and inclusive asymmetries are corrected
for detector and acceptance effects. Corrections are applied using bin-by-bin corrections for
\Acll{} and fully bayesian unfolding for \Actt. 
The distributions of lepton \deta and \ttbar \dy{} after the detector smearing
corrections are provided for the \emu{} channel. Good agreement 
between the corrected values and predictions of the Monte Carlo generator models is observed in these distributions. The combined values of lepton-based inclusive asymmetry \Acll and \ttbar\ inclusive asymmetry \Actt are measured to be 
 $\Acll = \Acllcomb$ and $\Actt = \Acttcomb$.
The measured values are in agreement with previous LHC measurements and with the
Standard Model prediction~\cite{Bernreuther:2012sx}: 
$\Acll = \Aclltheory$ and $\Actt = \Actttheory$.
The measurements are limited by statistical uncertainties. The predictions of benchmark light and heavy colour octet models with parameters selected such that the models are consistent with previous LHC and Tevatron data~\cite{Aguilar-Saavedra:2014nja} are found to be consistent with the measured asymmetries.

\section*{Acknowledgements}

We thank CERN for the very successful operation of the LHC, as well as the
support staff from our institutions without whom ATLAS could not be
operated efficiently.

We acknowledge the support of ANPCyT, Argentina; YerPhI, Armenia; ARC,
Australia; BMWFW and FWF, Austria; ANAS, Azerbaijan; SSTC, Belarus; CNPq and FAPESP,
Brazil; NSERC, NRC and CFI, Canada; CERN; CONICYT, Chile; CAS, MOST and NSFC,
China; COLCIENCIAS, Colombia; MSMT CR, MPO CR and VSC CR, Czech Republic;
DNRF, DNSRC and Lundbeck Foundation, Denmark; EPLANET, ERC and NSRF, European Union;
IN2P3-CNRS, CEA-DSM/IRFU, France; GNSF, Georgia; BMBF, DFG, HGF, MPG and AvH
Foundation, Germany; GSRT and NSRF, Greece; RGC, Hong Kong SAR, China; ISF, MINERVA, GIF, I-CORE and Benoziyo Center, Israel; INFN, Italy; MEXT and JSPS, Japan; CNRST, Morocco; FOM and NWO, Netherlands; BRF and RCN, Norway; MNiSW and NCN, Poland; GRICES and FCT, Portugal; MNE/IFA, Romania; MES of Russia and NRC KI, Russian Federation; JINR; MSTD,
Serbia; MSSR, Slovakia; ARRS and MIZ\v{S}, Slovenia; DST/NRF, South Africa;
MINECO, Spain; SRC and Wallenberg Foundation, Sweden; SER, SNSF and Cantons of
Bern and Geneva, Switzerland; NSC, Taiwan; TAEK, Turkey; STFC, the Royal
Society and Leverhulme Trust, United Kingdom; DOE and NSF, United States of
America.

The crucial computing support from all WLCG partners is acknowledged
gratefully, in particular from CERN and the ATLAS Tier-1 facilities at
TRIUMF (Canada), NDGF (Denmark, Norway, Sweden), CC-IN2P3 (France),
KIT/GridKA (Germany), INFN-CNAF (Italy), NL-T1 (Netherlands), PIC (Spain),
ASGC (Taiwan), RAL (UK) and BNL (USA) and in the Tier-2 facilities
worldwide.

\clearpage

\bibliographystyle{myamsplain}
\bibliography{JHEP_PR_dilasym}

\providecommand{\bysame}{\leavevmode\hbox to3em{\hrulefill}\thinspace}
\providecommand{\MR}{\relax\ifhmode\unskip\space\fi MR }
\providecommand{\MRhref}[2]{%
  \href{http://www.ams.org/mathscinet-getitem?mr=#1}{#2}
}
\providecommand{\href}[2]{#2}
\begin{thebibliography}{10}

\bibitem{TOP1}
{F. Abe et al., CDF Collaboration}, \emph{{Observation of top quark production
  in $p\bar{p}$ collisions}}, \emph{Phys.~Rev.~Lett.} \textbf{74} (1995)~2626
  [arXiv:hep-ex/9503002].

\bibitem{TOP2}
{A. Abachi et al., D0 Collaboration}, \emph{{Observation of the top quark}},
  \emph{Phys.~Rev.~Lett.} \textbf{74} (1995)~2632 [arXiv:hep-ex/9503003].

\bibitem{Frederix:2007gi}
{R. Frederix and F. Maltoni}, \emph{{Top pair invariant mass distribution: A
  Window on new physics}}, \emph{JHEP} \textbf{01} (2009)~047
  [arXiv:0712.2355].

\bibitem{Barger:2006hm}
{V. Barger, T. Han and D.~G.~E. Walker}, \emph{{Top Quark Pairs at High
  Invariant Mass: A Model-Independent Discriminator of New Physics at the
  LHC}}, \emph{Phys.~Rev.~Lett.} \textbf{100} (2008)~031801
  [arXiv:hep-ph/0612016].

\bibitem{AguilarSaavedra:2010zi}
{J.~A.~Aguilar-Saavedra}, \emph{{Effective four-fermion operators in top
  physics: A Roadmap}}, \emph{Nucl.~Phys.} \textbf{B~843} (2011)~638
  [arXiv:1008.3562].

\bibitem{Zhang:2010dr}
{C.~Zhang and S.~Willenbrock}, \emph{{Effective-Field-Theory Approach to
  Top-Quark Production and Decay}}, \emph{Phys.~Rev.} \textbf{D~83}
  (2011)~034006 [arXiv:1008.3869].

\bibitem{Hill:1994hp}
{C.~T. Hill}, \emph{{Topcolor assisted technicolor}}, \emph{Phys.~Lett.}
  \textbf{B~345} (1995)~483 [arXiv:hep-ph/9411426].

\bibitem{Cao:2004wd}
{J. Cao, G. Liu and J.~M. Yang}, \emph{{Probing topcolor-assisted technicolor
  from like-sign top pair production at CERN LHC}}, \emph{Phys.~Rev.}
  \textbf{D~70} (2004)~114035 [arXiv:hep-ph/0409334].

\bibitem{Aguilar-Saavedra:2014nja}
{J.~A. Aguilar-Saavedra}, \emph{{Portrait of a colour octet}}, \emph{JHEP}
  \textbf{08} (2014)~172 [arXiv:1405.5826].

\bibitem{Bernreuther:2012sx}
{W. Bernreuther and Z.-G. Si}, \emph{{Top quark and leptonic charge asymmetries
  for the Tevatron and LHC}}, \emph{Phys.~Rev.} \textbf{D~86} (2012)~034026
  [arXiv:1205.6580].

\bibitem{Czakon:2014xsa}
M.~Czakon, P.~Fiedler, and A.~Mitov, \emph{{Resolving the Tevatron top quark
  forward-backward asymmetry puzzle}} [arXiv:1411.3007].

\bibitem{Wang:2014sua}
{J. S. Brodsky et al.}, \emph{{Application of the Principle of Maximum
  Conformality to the Top-Quark Charge Asymmetry at the LHC}},
  \emph{Phys.~Rev.} \textbf{D~90} (2014)~114034 [arXiv:1410.1607].

\bibitem{Frampton:1987dn}
{P.~H. Frampton and S.~L. Glashow}, \emph{{Chiral Color: An Alternative to the
  Standard Model}}, \emph{Phys.~Lett.} \textbf{B~190} (1987)~157.

\bibitem{Ferrario:2008wm}
{P. Ferrario and G. Rodrigo}, \emph{{Massive color-octet bosons and the charge
  asymmetries of top quarks at hadron colliders}}, \emph{Phys.~Rev.}
  \textbf{D~78} (2008)~094018 [arXiv:0809.3354].

\bibitem{AguilarSaavedra:2011vw}
{J.~A. Aguilar-Saavedra and M. Perez-Victoria}, \emph{{Probing the Tevatron
  \ttbar{} asymmetry at LHC}}, \emph{JHEP} \textbf{05} (2011)~034
  [arXiv:1103.2765].

\bibitem{Falkowski:2012cu}
{A. Falkowski et al.}, \emph{{Data driving the top quark forward-backward
  asymmetry with a lepton-based handle}}, \emph{Phys.~Rev.} \textbf{D~87}
  (2013)~034039 [arXiv:1212.4003].

\bibitem{Aguilar-Saavedra:2014kpa}
{J.~A. Aguilar-Saavedra et al.}, \emph{{Asymmetries in top quark pair
  production}},  (2014) [arXiv:1406.1798].

\bibitem{Aad:2013cea}
{ATLAS Collaboration}, \emph{{Measurement of the top quark pair production
  charge asymmetry in proton-proton collisions at $\sqrt{s}$ = 7 TeV using the
  ATLAS detector}}, \emph{JHEP} \textbf{02} (2014)~107 [arXiv:1311.6724].

\bibitem{Chatrchyan:2012xv}
{CMS Collaboration}, \emph{{Inclusive and differential measurements of the $t
  \bar{t}$ charge asymmetry in proton-proton collisions at 7 TeV}},
  \emph{Phys.~Lett.} \textbf{B~717} (2012)~129 [arXiv:1207.0065].

\bibitem{Chatrchyan:2014yta}
{CMS Collaboration}, \emph{{Measurements of the $t\bar{t}$ charge asymmetry
  using the dilepton decay channel in pp collisions at $\sqrt{s} =$ 7 TeV}},
  \emph{JHEP} \textbf{04} (2014)~191 [arXiv:1402.3803].

\bibitem{ATLAS-CONF-2014-012}
{ATLAS and CMS Collaborations}, \emph{{Combination of ATLAS and CMS $t\bar{t}$
  charge asymmetry measurements using LHC proton-proton collisions at
  $\sqrt{s}=7$~\tev}},
  {ATLAS-CONF-2014-012,~CMS-PAS-TOP-14-006,~\href{https://cds.cern.ch/record/1%
670535}{https://cds.cern.ch/record/1670535}} (2014).

\bibitem{CDF1}
{T. Aaltonen et al., CDF Collaboration}, \emph{{Evidence for a Mass Dependent
  Forward-Backward Asymmetry in Top Quark Pair Production}}, \emph{Phys.~Rev.}
  \textbf{D~83} (2011)~112003 [arXiv:1101.0034].

\bibitem{D01}
{V.~M. Abazov et al., D0 Collaboration}, \emph{{Forward-backward asymmetry in
  top quark-antiquark production}}, \emph{Phys.~Rev.} \textbf{D~84}
  (2011)~112005 [arXiv:1107.4995].

\bibitem{CDF2}
{T. Aaltonen et al., CDF Collaboration}, \emph{{Measurement of the top quark
  forward-backward production asymmetry and its dependence on event kinematic
  properties}}, \emph{Phys.~Rev.} \textbf{D~87} (2013)~092002
  [arXiv:1211.1003].

\bibitem{PhysRevD.88.072003}
{T. Aaltonen et al., CDF Collaboration}, \emph{{Measurement of the leptonic
  asymmetry in $t \bar t$ events produced in $p \bar p$ collisions at
  $\sqrt{s}=$ 1.96 TeV}}, \emph{Phys.~Rev.} \textbf{88} (2013)~072003
  [arXiv:1308.1120].

\bibitem{Aaltonen:2014eva}
{T. Aaltonen et al., CDF Collaboration}, \emph{{Measurement of the inclusive
  leptonic asymmetry in top-quark pairs that decay to two charged leptons at
  CDF}}, \emph{Phys.~Rev.~Lett.} \textbf{113} (2014)~042001 [arXiv:1404.3698].

\bibitem{Abazov:2014oea}
{V.~M. Abazov et al., D0 Collaboration}, \emph{{Measurement of the
  forward-backward asymmetry in the distribution of leptons in $t\bar{t}$
  events in the lepton+jets channel}}, \emph{Phys.~Rev.} \textbf{D~90}
  (2014)~072001 [arXiv:1403.1294].

\bibitem{D02}
{V.~M. Abazov et al., D0 Collaboration}, \emph{{Measurement of the
  forward-backward asymmetry in top quark-antiquark production in \ppbar{}
  collisions using the lepton+jets channel}}, \emph{Phys.~Rev.} \textbf{D~90}
  (2014)~072011 [arXiv:1405.0421].

\bibitem{atlas}
{ATLAS Collaboration}, \emph{{The ATLAS Experiment at the CERN Large Hadron
  Collider}}, \emph{JINST} \textbf{3} (2008)~S08003.

\bibitem{FRI-0701}
{S. Frixione, P. Nason and C. Oleari}, \emph{Matching NLO QCD computations with
  Parton Shower simulations: the POWHEG method}, \emph{JHEP} \textbf{11}
  (2007)~070 [arXiv:0709.2092].

\bibitem{Nason:2004rx}
{P. Nason}, \emph{{A New method for combining NLO QCD with shower Monte Carlo
  algorithms}}, \emph{JHEP} \textbf{11} (2004)~040.

\bibitem{Frixione:2007nw}
{S. Frixione, G. Ridolfi and P. Nason}, \emph{{A Positive-weight
  next-to-leading-order Monte Carlo for heavy flavour hadroproduction}},
  \emph{JHEP} \textbf{09} (2007)~126 [arXiv:0707.3088].

\bibitem{Lai:2010vv}
{H.~L. Lai et al.}, \emph{{New parton distributions for collider physics}},
  \emph{Phys.~Rev.} \textbf{D~82} (2010)~074024 [arXiv:1007.2241].

\bibitem{SJO-0601}
{T. Sj\"ostrand, S. Mrenna and P.~Z. Skands}, \emph{PYTHIA 6.4 Physics and
  Manual}, \emph{JHEP} \textbf{05} (2006)~026 [arXiv:hep-ph/0603175].

\bibitem{cteq6}
{J. Pumplin et al.}, \emph{{New generation of parton distributions with
  uncertainties from global QCD analysis}}, \emph{JHEP} \textbf{07} (2002)~012
  [arXiv:hep-ph/0201195].

\bibitem{PhysRevD.82.074018}
{P.~Z. Skands}, \emph{{Tuning Monte Carlo Generators: The Perugia Tunes}},
  \emph{Phys.~Rev.} \textbf{D~82} (2010)~074018 [arXiv:1005.3457].

\bibitem{Cacciari:2011hy}
{M. Cacciari et al.}, \emph{{Top-pair production at hadron colliders with
  next-to-next-to-leading logarithmic soft-gluon resummation}},
  \emph{Phys.~Lett.} \textbf{B~710} (2012)~612 [arXiv:1111.5869].

\bibitem{Baernreuther:2012ws}
{P. Baernreuther, M. Czakon and A. Mitov}, \emph{{Percent Level Precision
  Physics at the Tevatron: First Genuine NNLO QCD Corrections to $q \bar{q} \to
  t \bar{t} + X$}}, \emph{Phys.~Rev.~Lett.} \textbf{109} (2012)~132001
  [arXiv:1204.5201].

\bibitem{Czakon:2012zr}
{M. Czakon and A. Mitov}, \emph{{NNLO corrections to top-pair production at
  hadron colliders: the all-fermionic scattering channels}}, \emph{JHEP}
  \textbf{12} (2012)~054 [arXiv:1207.0236].

\bibitem{Czakon:2012pz}
{M. Czakon and A. Mitov}, \emph{{NNLO corrections to top pair production at
  hadron colliders: the quark-gluon reaction}}, \emph{JHEP} \textbf{01}
  (2013)~080 [arXiv:1210.6832].

\bibitem{Czakon:2013goa}
{M. Czakon, P. Fiedler and A. Mitov}, \emph{{Total Top-Quark Pair-Production
  Cross Section at Hadron Colliders Through $O(\alpha_{S}^{4})$}},
  \emph{Phys.~Rev.~Lett.} \textbf{110} (2013)~252004 [arXiv:1303.6254].

\bibitem{Beneke:2011mq}
{M. Beneke et al.}, \emph{{Hadronic top-quark pair production with NNLL
  threshold resummation}}, \emph{Nucl.~Phys.} \textbf{B~855} (2012)~695
  [arXiv:1109.1536].

\bibitem{Czakon:2011xx}
{M. Czakon and A. Mitov}, \emph{{Top++: A Program for the Calculation of the
  Top-Pair Cross-Section at Hadron Colliders}}, \emph{Comput.~Phys.~Commun.}
  \textbf{185} (2014)~2930 [arXiv:1112.5675].

\bibitem{Botje:2011sn}
{M~Botje et al.}, \emph{{The PDF4LHC Working Group Interim Recommendations}},
  (2011) [arXiv:1101.0538].

\bibitem{Martin:2009iq}
{A.~D. Martin et al.}, \emph{{Parton distributions for the LHC}},
  \emph{Eur.~Phys.~J.} \textbf{C~63} (2009)~189 [arXiv:0901.0002].

\bibitem{Martin:2009bu}
{A.~D. Martin et al.}, \emph{{Uncertainties on $\alpha_{S}$ in global PDF
  analyses and implications for predicted hadronic cross sections}},
  \emph{Eur.~Phys.~J.} \textbf{C~64} (2009)~653 [arXiv:0905.3531].

\bibitem{Gao:2013xoa}
{J. Gao et al.}, \emph{{The CT10 NNLO Global Analysis of QCD}},
  \emph{Phys.~Rev.} \textbf{D~89} (2014)~033009 [arXiv:1302.6246].

\bibitem{Ball:2012cx}
{R.~D. Ball et al.}, \emph{{Parton distributions with LHC data}},
  \emph{Nucl.~Phys.} \textbf{B~867} (2013)~244 [arXiv:1207.1303].

\bibitem{Aliev:2010zk}
{M. Aliev et al.}, \emph{{HATHOR: HAdronic Top and Heavy quarks crOss section
  calculatoR}}, \emph{Comput.~Phys.~Commun.} \textbf{182} (2011)~1034
  [arXiv:1007.1327].

\bibitem{FRI-0201}
{S. Frixione and B.~R. Webber}, \emph{{Matching NLO QCD computations and parton
  shower simulations}}, \emph{JHEP} \textbf{06} (2002)~029
  [arXiv:hep-ph/0204244].

\bibitem{Frixione:2008yi}
{S. Frixione et al.}, \emph{{Single-top hadroproduction in association with a W
  boson}}, \emph{JHEP} \textbf{07} (2008)~029 [arXiv:0805.3067].

\bibitem{Marchesini:1991ch}
{G. Marchesini et al.}, \emph{{HERWIG 5.1 - a Monte Carlo event generator for
  simulating hadron emission reactions with interfering gluons}},
  \emph{Comput.~Phys.~Commun.} \textbf{67} (1992)~465.

\bibitem{COR-0001}
{G. Corcella et al.}, \emph{{HERWIG 6: An Event generator for hadron emission
  reactions with interfering gluons (including supersymmetric processes)}},
  \emph{JHEP} \textbf{01} (2001)~010 [arXiv:hep-ph/0011363].

\bibitem{JButterworth:1996zw}
{J.~M. Butterworth et al.}, \emph{{Multiparton interactions in photoproduction
  at HERA}}, \emph{Z.~Phys.} \textbf{C~72} (1996)~637 [arXiv:hep-ph/9601371].

\bibitem{PUB-2011-008}
{ATLAS Collaboration}, \emph{{New ATLAS event generator tunes to 2010 data}},
  {ATL-PHYS-PUB-2011-008,~\href{http://cdsweb.cern.ch/record/1345343}{http://c%
dsweb.cern.ch/record/1345343}} (2011).

\bibitem{MAN-0301}
{M.~L. Mangano et al.}, \emph{ALPGEN, a generator for hard multiparton
  processes in hadronic collisions}, \emph{JHEP} \textbf{07} (2003)~001
  [arXiv:hep-ex/0206293].

\bibitem{Kidonakis:2010ux}
{N. Kidonakis}, \emph{{Two-loop soft anomalous dimensions for single top quark
  associated production with a W- or H-}}, \emph{Phys.~Rev.} \textbf{D~82}
  (2010)~054018 [arXiv:1005.4451].

\bibitem{Campbell:1999ah}
{J.~M. Campbell and R.~K. Ellis}, \emph{{An Update on vector boson pair
  production at hadron colliders}}, \emph{Phys.~Rev.} \textbf{D~60}
  (1999)~113006 [arXiv:hep-ph/9905386].

\bibitem{Anastasiou:2003ds}
{C.~Anastasiou et al.}, \emph{{High precision QCD at hadron colliders:
  Electroweak gauge boson rapidity distributions at NNLO}}, \emph{Phys.~Rev.}
  \textbf{D~69} (2004)~094008 [arXiv:hep-ph/0312266].

\bibitem{Hamberg:1990np}
{R.~Hamberg, W.~L. van~Neerven and T. Matsuura}, \emph{{A Complete calculation
  of the order ${\alpha}_s^{2}$ correction to the Drell-Yan $K$ factor}},
  \emph{Nucl.~Phys.} \textbf{B~359} (1991)~343, {Erratum-ibid. {\bf B 644}
  (2002) 403}.

\bibitem{atlasfullsim}
{ATLAS Collaboration}, \emph{{The ATLAS Simulation Infrastructure}},
  \emph{Eur.~Phys.~J.} \textbf{C~70} (2010)~823 [arXiv:1005.4568].

\bibitem{Agostinelli:2002hh}
{S. Agostinelli et al.}, \emph{{GEANT4: A simulation toolkit}},
  \emph{Nucl.~Instrum.~Meth.} \textbf{A~506} (2003)~250.

\bibitem{atlasfastsim}
{ATLAS Collaboration}, \emph{{The simulation principle and performance of the
  ATLAS fast calorimeter simulation FastCaloSim}},
  {ATL-PHYS-PUB-2010-013,~\href{http://cdsweb.cern.ch/record/1300517}{http://c%
dsweb.cern.ch/record/1300517}} (2010).

\bibitem{Aad:2013ucp}
{ATLAS Collaboration}, \emph{{Improved luminosity determination in $pp$
  collisions at $\sqrt{s}$ = 7 TeV using the ATLAS detector at the LHC}},
  \emph{Eur.~Phys.~J.} \textbf{C~73} (2013)~2518 [arXiv:1302.4393].

\bibitem{Aad:2014fxa}
{ATLAS Collaboration}, \emph{{Electron reconstruction and identification
  efficiency measurements with the ATLAS detector using the 2011 LHC
  proton-proton collision data}}, \emph{Eur.~Phys.~J.} \textbf{C~74}
  (2014)~2941 [arXiv:1404.2240].

\bibitem{Aad:2014zya}
{ATLAS Collaboration}, \emph{{Measurement of the muon reconstruction
  performance of the ATLAS detector using 2011 and 2012 LHC proton-proton
  collision data}}, \emph{Eur.~Phys.~J.} \textbf{C~74} (2014)~3130
  [arXiv:1407.3935].

\bibitem{Cacciari:2008gp}
{M. Cacciari, G.~P. Salam and G. Soyez}, \emph{{The anti-kt jet clustering
  algorithm}}, \emph{JHEP} \textbf{04} (2008)~063 [arXiv:0802.1189].

\bibitem{atlastopo}
{ATLAS Collaboration}, \emph{{Calorimeter Clustering Algorithms: Description
  and Performance}}, {ATL-LARG-PUB-2008-002,
  \href{https://cds.cern.ch/record/1099735}{https://cds.cern.ch/record/1099735%
}} (2010).

\bibitem{Aad:2014bia}
{ATLAS Collaboration}, \emph{{Jet energy measurement and its systematic
  uncertainty in proton-proton collisions at $\sqrt{s}=7$ TeV with the ATLAS
  detector}}, \emph{Eur.~Phys.~J.} \textbf{C~75} (2015)~17 [arXiv:1406.0076].

\bibitem{Aad:2012re}
{ATLAS Collaboration}, \emph{{Performance of Missing Transverse Momentum
  Reconstruction in Proton-Proton Collisions at 7 TeV with ATLAS}},
  \emph{Eur.~Phys.~J.} \textbf{C~72} (2012)~1844 [arXiv:1108.5602].

\bibitem{Aad:2010ey}
{ATLAS Collaboration}, \emph{{Measurement of the top quark-pair production
  cross section with ATLAS in pp collisions at $\sqrt{s}=7$ TeV}},
  \emph{Eur.~Phys.~J.} \textbf{C~71} (2011)~1577 [arXiv:1012.1792].

\bibitem{Abbott:1997fv}
{B. Abbott et al., D0 Collaboration}, \emph{{Measurement of the top quark mass
  using dilepton events}}, \emph{Phys.~Rev.~Lett.} \textbf{80} (1998)~2063
  [arXiv:hep-ex/9706014].

\bibitem{Fbu2012arXiv1201.4612C}
{G. Choudalakis}, \emph{{Fully Bayesian Unfolding}} [arXiv:1201.4612].

\bibitem{Aad:2012ag}
{ATLAS Collaboration}, \emph{{Jet energy resolution in proton-proton collisions
  at $\sqrt{s}=7$ TeV recorded in 2010 with the ATLAS detector}},
  \emph{Eur.~Phys.~J.} \textbf{C~73} (2013)~2306 [arXiv:1210.6210].

\bibitem{Lyons:1988rp}
{L. Lyons, D. Gibaut and P. Clifford}, \emph{{How to Combine Correlated
  Estimates of a Single Physical Quantity}}, \emph{Nucl.~Instrum.~Meth.}
  \textbf{A~270} (1988)~110.

\bibitem{Valassi:2003mu}
{A. Valassi}, \emph{{Combining correlated measurements of several different
  physical quantities}}, \emph{Nucl.~Instrum.~Meth.} \textbf{A~500} (2003)~391.

\end{thebibliography}

\newpage
\appendix

\section{Additional tables}
Additional information about the normalized differential cross-sections in the \emu{} channel are provided in this appendix.

The detail of the systematic uncertainties in each bin of the distributions are reported in
\tabsAndTabRef{tab:comb_syst_leptons}{tab:comb_syst_ttbar}.

\begin{table}[ht]
\begin{center}
\begingroup\fontsize{7pt}{11pt}\selectfont
\begin{tabular*}{1.1\textwidth}{@{\extracolsep{\fill} } l @{}c@{}c@{}c@{}c@{}c@{}c@{}c}
Bin  of \deta & $[-3.,-2.]$ & $[-2.,-1.67]$ & $[-1.67,-1.33]$ & $[-1.33,-1.]$ & $[-1.,-0.67]$ & $[-0.67,-0.33]$ & $[-0.33,0.]$  \\
\hline

Central value         & 0.0440 & 0.106 & 0.126 & 0.164 & 0.245  & 0.314 & 0.400   \\
Stat.                 & 0.0077 & 0.011 & 0.011 & 0.012 & 0.013  & 0.015 & 0.016   \\
\hline
Lepton reconstruction & 0.0023 & 0.003 & 0.005 & 0.003 & 0.003  & 0.005 & 0.003   \\
Jet reconstruction    & 0.0005 & 0.001 & -     & 0.002 & -      & 0.002 & 0.001   \\
\met{}                & -      & -     & -     & -     & -      & -     & -       \\
Signal modelling      & n/r    & n/r   & n/r   & n/r   & n/r    & n/r   & n/r   \\
PDF                   & n/r    & n/r   & n/r   & n/r   & n/r    & n/r   & n/r   \\
Background            & -      & 0.001 & 0.001 & -     & 0.001  & 0.002 & 0.002   \\
NP and fake           & 0.0005 & 0.002 & 0.001 & -     & 0.001  & 0.004 & 0.001   \\
\hline
Total systematics     & 0.0025 & 0.004 & 0.005 & 0.004 & 0.004 & 0.007 & 0.004   \\
Total uncertainty     & 0.0081 & 0.011 & 0.012 & 0.013 & 0.014 & 0.016 & 0.017   \\

\end{tabular*}
\endgroup

\vspace{0.2cm}
\begingroup\fontsize{7pt}{11pt}\selectfont
\begin{tabular*}{1.1\textwidth}{@{\extracolsep{\fill} } l @{}c@{}c@{}c@{}c@{}c@{}c@{}c}
Bin of \deta & $[0.,0.33]$ & $[0.33,0.67]$ & $[0.67,1.]$ & $[1.,1.33]$ & $[1.33,1.67]$ & $[1.67,2.]$ & $[2.,3.]$ \\
\hline

Central value         & 0.392 & 0.349 & 0.244 & 0.209 & 0.129 & 0.0815 & 0.0361   \\
Stat.                 & 0.016 & 0.015 & 0.013 & 0.013 & 0.011 & 0.0093 & 0.0076   \\
\hline
Lepton reconstruction & 0.003 & 0.009 & 0.009 & 0.003 & 0.003 & 0.0019 & 0.0032   \\
Jet reconstruction    & 0.001 & 0.001 & 0.002 & 0.002 & -     & 0.0012 & 0.0003   \\
\met{}                & -     & -     & -     & -     & -     & -      & -        \\
Signal modelling      & n/r   & n/r   & n/r   & n/r   & n/r   & n/r    & n/r   \\
PDF                   & n/r   & n/r   & n/r   & n/r   & n/r   & n/r    & n/r   \\
Background            & 0.002 & -     & -     & 0.001 & 0.001 & 0.0009 & -     \\
NP and fake           & 0.001 & 0.001 & 0.002 & 0.003 & 0.001 & 0.0012 & 0.0013   \\
\hline
Total systematics     & 0.004 & 0.009 & 0.010 & 0.005 & 0.003 & 0.0028 & 0.0035   \\
Total uncertainty     & 0.017 & 0.018 & 0.017 & 0.014 & 0.011 & 0.0097 & 0.0084  \\

\end{tabular*}
\endgroup

\caption{Systematic uncertainties in each bin of the \deta distribution in the \emu{} channel. Hyphens are used when the uncertainties are lower than 0.0005. The signal modelling and the PDF uncertainty, (labeled as n/r) are limited by the statistical fluctuations in the simulated samples and are thus not reported in the table.}
\label{tab:comb_syst_leptons}
\end{center}
\end{table}

\begin{table}[ht]
\begin{center}
{\footnotesize 

\begin{tabular}{lcccc}
Bin of \dy                       & $[-5.00,-0.71]$ & $[-0.71,0.00]$ & $[0.00,0.71]$ & $[0.71,5.00]$ \\
\hline

Central value          & 0.0435 & 0.437 & 0.420 & 0.0470   \\
Stat.                 & 0.0029 & 0.016 & 0.025 & 0.0032   \\
\hline
Lepton reconstruction & 0.0009 & 0.006 & 0.010 & 0.0016   \\
Jet reconstruction   & 0.0007 & 0.011 & 0.009 & 0.0012   \\
\met{}              & 0.0002 & 0.003 & 0.007 & 0.0009   \\
Signal modelling     & n/r & n/r & n/r & n/r   \\
PDF                  & n/r & n/r & n/r & n/r   \\
Background           & 0.0003 & 0.001 & 0.001 & 0.0002   \\
NP and fake          & 0.0011 & 0.001 & 0.004 & 0.0007   \\
\hline
Total systematics    & 0.0017 & 0.013 & 0.015 & 0.0024   \\
Total uncertainty    & 0.0034 & 0.021 & 0.029 & 0.0040  \\        


\hline
\end{tabular}

}
\caption{Systematic uncertainties in each bin of the \dy distribution in the \emu{} channel. The signal modelling and the PDF uncertainty, (labeled as n/r) are limited by the statistical fluctuations in the simulated samples and are thus not reported in the table.}
\label{tab:comb_syst_ttbar}
\end{center}
\end{table}

The statistical correlations between the different bins of each distribution are reported in \tabsAndTabRef{tab:corr_leptons}{tab:corr_ttbar}. They were estimated using bootstrapping.

\begin{table}[ht]
\begin{center}
\begingroup\fontsize{7pt}{11pt}\selectfont 
\begin{tabular}{c|cccccccccccccc}
 & 1 & 2 & 3 & 4 & 5 & 6 & 7 & 8 & 9 & 10 & 11 & 12 & 13 & 14 \\
\hline   
  1 & $+$1.00 & $-$0.49 & $-$0.02 & $-$0.03 & $-$0.02 & $+$0.01 & $-$0.03 & $+$0.02 & $+$0.06 & $-$0.06 & $-$0.01 & $+$0.03 & $-$0.02 & $+$0.03  \\ 
  2 & $-$0.49 & $+$1.00 & $-$0.50 & $-$0.04 & $+$0.02 & $+$0.00 & $+$0.00 & $+$0.01 & $-$0.04 & $+$0.05 & $-$0.02 & $-$0.01 & $+$0.01 & $-$0.03   \\ 
  3 & $-$0.02 & $-$0.50 & $+$1.00 & $-$0.47 & $+$0.00 & $+$0.02 & $-$0.03 & $-$0.02 & $-$0.03 & $+$0.03 & $+$0.03 & $+$0.01 & $-$0.02 & $-$0.00   \\ 
  4 & $-$0.03 & $-$0.04 & $-$0.47 & $+$1.00 & $-$0.49 & $-$0.04 & $-$0.00 & $+$0.04 & $+$0.03 & $-$0.05 & $+$0.03 & $-$0.03 & $+$0.03 & $-$0.01   \\ 
  5 & $-$0.02 & $+$0.02 & $+$0.00 & $-$0.49 & $+$1.00 & $-$0.52 & $+$0.06 & $-$0.04 & $-$0.01 & $+$0.03 & $-$0.05 & $+$0.00 & $-$0.04 & $+$0.06   \\ 
  6 & $+$0.01 & $+$0.00 & $+$0.02 & $-$0.04 & $-$0.52 & $+$1.00 & $-$0.54 & $+$0.04 & $-$0.02 & $+$0.01 & $+$0.01 & $-$0.00 & $+$0.02 & $-$0.01   \\ 
  7 & $-$0.03 & $+$0.00 & $-$0.03 & $-$0.00 & $+$0.06 & $-$0.54 & $+$1.00 & $-$0.54 & $-$0.00 & $-$0.00 & $-$0.00 & $+$0.04 & $-$0.02 & $+$0.01   \\ 
  8 & $+$0.02 & $+$0.01 & $-$0.02 & $+$0.04 & $-$0.04 & $+$0.04 & $-$0.54 & $+$1.00 & $-$0.53 & $+$0.02 & $+$0.02 & $-$0.05 & $+$0.00 & $-$0.02   \\ 
  9 & $+$0.06 & $-$0.04 & $-$0.03 & $+$0.03 & $-$0.01 & $-$0.02 & $-$0.00 & $-$0.53 & $+$1.00 & $-$0.52 & $+$0.01 & $+$0.00 & $+$0.03 & $-$0.02   \\ 
  10 & $-$0.06 & $+$0.05 & $+$0.03 & $-$0.05 & $+$0.03 & $+$0.01 & $-$0.00 & $+$0.02 & $-$0.52 & $+$1.00 & $-$0.52 & $+$0.02 & $-$0.02 & $+$0.01   \\ 
  11 & $-$0.01 & $-$0.02 & $+$0.03 & $+$0.03 & $-$0.05 & $+$0.01 & $-$0.00 & $+$0.02 & $+$0.01 & $-$0.52 & $+$1.00 & $-$0.49 & $+$0.03 & $-$0.03   \\ 
  12 & $+$0.03 & $-$0.01 & $+$0.01 & $-$0.03 & $+$0.00 & $-$0.00 & $+$0.04 & $-$0.05 & $+$0.00 & $+$0.02 & $-$0.49 & $+$1.00 & $-$0.56 & $+$0.02   \\ 
  13 & $-$0.02 & $+$0.01 & $-$0.02 & $+$0.03 & $-$0.04 & $+$0.02 & $-$0.02 & $+$0.00 & $+$0.03 & $-$0.02 & $+$0.03 & $-$0.56 & $+$1.00 & $-$0.54   \\ 
  14 & $+$0.03 & $-$0.03 & $-$0.00 & $-$0.01 & $+$0.06 & $-$0.01 & $+$0.01 & $-$0.02 & $-$0.02 & $+$0.01 & $-$0.03 & $+$0.02 & $-$0.54 & $+$1.00   \\ 
\end{tabular}
\endgroup

\caption{Statistical bin-bin correlations within the \deta distribution in the \emu{} channel. The bin numbers are used instead of the bin boundaries. 
  Bin 1 corresponds to $[-3.00, -2.00]$ and bin 14 to $[2.00, 3.00]$.}
\label{tab:corr_leptons}
\end{center}
\end{table}

\begin{table}[ht]
\begin{center}
{\footnotesize
\begin{tabular}{c|cccc}
 & 1 & 2 & 3 & 4 \\
\hline   
   1 &     $+$1.00 & -0.63 & $+$0.38 & -0.38 \\
   2 &    -0.63 & $+$1.00 & -0.79 & $+$0.37 \\
   3 &    $+$0.38 & -0.79 & $+$1.00 & -0.61 \\
   4 &    -0.38 & $+$0.37 & -0.61 & $+$1.00 \\
\end{tabular}

}
\caption{Statistical bin-bin correlations within the \dy distribution in the \emu{} channel. The bin numbers are used instead of the bin boundaries.
  Bin 1 corresponds to $[-5.00,-0.71]$ and bin 4 to $[0.71, 5.00]$.}
\label{tab:corr_ttbar}
\end{center}
\end{table}

\label{app:Appendix}

\onecolumn
\clearpage 
\begin{flushleft}
{\Large The ATLAS Collaboration}

\bigskip

G.~Aad$^{\rm 85}$,
B.~Abbott$^{\rm 113}$,
J.~Abdallah$^{\rm 152}$,
S.~Abdel~Khalek$^{\rm 117}$,
O.~Abdinov$^{\rm 11}$,
R.~Aben$^{\rm 107}$,
B.~Abi$^{\rm 114}$,
M.~Abolins$^{\rm 90}$,
O.S.~AbouZeid$^{\rm 159}$,
H.~Abramowicz$^{\rm 154}$
% ...
\bigskip
\\
% ...
$^{11}$ Institute of Physics, Azerbaijan Academy of Sciences, Baku, Azerbaijan\\
% ...
$^{85}$ CPPM, Aix-Marseille Universit{\'e} and CNRS/IN2P3, Marseille, France\\
% ...
$^{90}$ Department of Physics and Astronomy, Michigan State University, East Lansing MI, United States of America\\
% ...
$^{107}$ Nikhef National Institute for Subatomic Physics and University of Amsterdam, Amsterdam, Netherlands\\
% ...
$^{113}$ Homer L. Dodge Department of Physics and Astronomy, University of Oklahoma, Norman OK, United States of America\\
$^{114}$ Department of Physics, Oklahoma State University, Stillwater OK, United States of America\\
% ...
$^{117}$ LAL, Universit{\'e} Paris-Sud and CNRS/IN2P3, Orsay, France\\
% ...
$^{152}$ Institute of Physics, Academia Sinica, Taipei, Taiwan\\
% ...
$^{154}$ Raymond and Beverly Sackler School of Physics and Astronomy, Tel Aviv University, Tel Aviv, Israel\\
% ...
$^{159}$ Department of Physics, University of Toronto, Toronto ON, Canada\\
% ...
\end{flushleft}

\end{document}